# BALANCING TRANSPARENCY, EFFICIENCY, AND SECURITY IN PERVASIVE SYSTEMS

Mark Wenstrom, Eloisa Bentivegna*, and Ali R. Hurson

Department of Computer Science and Engineering
Penn State University
University Park, PA 16802
{wenstrom, hurson}@cse.psu.edu

*Department of Physics
Penn State University
University Park, PA 16802
bentiveg@phys.psu.edu

## Abstract

This chapter will survey pervasive computing with a look at how its constraint for transparency affects issues of resource management and security. The goal of pervasive computing is to render computing transparent, such that computing resources are ubiquitously offered to the user and services are proactively performed for a user without his or her intervention. The task of integrating computing infrastructure into everyday life without making it excessively invasive brings about tradeoffs between flexibility and robustness, efficiency and effectiveness, as well as autonomy and reliability. While not the primary goal of pervasive computing, efficiency in resource management should be considered in order to best utilize a limited set of resources (bandwidth, computing, etc.) as to avoid congestion and avoid creating a visible and distracting bottleneck in the eyes of the user. As solutions to efficiently manage the resources in a pervasive computing environment, three techniques will be examined: the distributed caching and sharing of data between mobile hosts, the broadcasting of services by public service providers, and the ability for mobile hosts to adaptively adjust the quality of offered services. Likewise, security is often an afterthought in many computing projects, though it should be of high consideration in a pervasive environment where users share public resources to operate on private data. Specifically, how can a user be authenticated in this environment with minimal or no user intervention? Solutions such as single sign-on via smartcards and biometrics will be examined to carry out authentication in a pervasive environment. As the feasibility of ubiquitous computing and its real potential for mass applications are still a matter of controversy, this chapter will look into the underlying issues of resource management and authentication to discover how these can be handled in a least invasive fashion. The discussion will be closed by an overview of the solutions proposed by current pervasive computing efforts, both in the area of generic platforms and for dedicated applications such as pervasive education and healthcare.

# Table of Contents

# List of Abbreviations

| | |
|---|---|
| FMR | false match rate |
| FNM | false nonmatch rate |
| FPS | frames per second |
| FR | read frequency |
| LRU | least recently used |
| MTBR | mean time between reads |
| MTBU | mean time between updates |
| P2P | peer-to-peer |
| PDA | personal digital assistant |
| PDF | probability density function |
| PM | probability that an object has been modified |
| PNM | probability that an object has not been modified |
| QoS | quality of service |
| SDS | secure discovery service |
| STDV | standard deviation |
| TGS | ticket granting service |
| TSP | traveling salesman problem |

# 1 Introduction

Pervasive computing explores the task of integrating technology into an environment, such that a multitude of computing devices are available to proactively perform services for each user, thereby lightening the user's workload. It has been pointed out [1] that pervasive systems constitute the third wave in computing, after the main frame era (one computer, many users) and the personal computer era (one computer, one user). Pervasive computing is the next natural step in order to set a single user in control of several computing elements. It should be noted that occasionally, the literature has used the term "ubiquitous computing" and "pervasive computing" interchangeably. In this article, however, we are making a distinction between the two: Pervasive refers to the invisibility and proactivity where the computer dissolves into the fabric of the surroundings while ubiquity refers to the availability. In another words, ubiquitous computing facilitates a better pervasive computing.

In a pervasive environment a user should always have access to computing resources, whether the user is mobile or stationary; thus it is assumed that each user is in control of a personal mobile device (e.g., a PDA or laptop). One's device can be used to connect to a multitude of resources: it can wirelessly tap into an access point for connectivity to the global Internet; similarly, the device can tap into a local access point for connectivity to a local network (e.g., a university or office network); on the other hand, the device may not be able to tap into a structured network, but may be able to join an ad hoc network of wireless devices in order to utilize those resources.

As of the year 2007, the aforementioned forms of connectivity and resource sharing are widely available. The field of mobile computing has provided some of the framework for pervasive computing, as mobile devices currently give users access to computing resource at all times. Not only does pervasive computing require the constant access to resources, but it also requires that technology be seamlessly and invisibly integrated into the lives of its users. Thus, the field of artificial intelligence must also be employed, such that one's mobile device can predict the desires of its user and can independently carry out services for this user. This implies that a mobile device must be aware of its surrounding context, and must be able to locate and call upon remote resources to carry out its user's intent. Consequently, the field of distributed computing must be employed to provide techniques to divide out computations to remote resources; this is especially important to a pervasive environment, as the environment consists of a variety of computing elements, ranging from powerful servers to resource-constrained mobile devices.

These areas of computing can be adapted, fused and ameliorated to achieve the goal of pervasive computing: balancing proactivity of services and transparency of operation in order to saturate an environment with computing agents that automate the trivial daily tasks of life (e.g., transferring one's lecture notes from a PDA to a workstation), leaving humans free to focus on the high-level tasks (e.g., delivering a lecture). In other words, the focus of one's action is intended to be the high-level task rather than the technology.

Notwithstanding the availability of the required technology, true pervasive computing environments have not been realized, as only prototypes and theoretical designs have been developed by the research community. A major open field is related to the fact that pervasive computing faces the delicate issue of which choices can be delegated to the system (in the form of local clients, neighboring peers or a central server) and which must be imperatively performed by the user. The goal of pervasive computing is obviously to maximize the former and minimize the latter. This often demands for smarter algorithms, architectures and technologies than are presently available. In order to create a system that proactively issues tasks, yet remains mostly transparent from the user, two broad-range issues, with ramifications and follow-ons need to be addressed:

- **The computing agents need to be able to predict the user's intent based on history and context-awareness**. Predicting a user's intent implies that the services performed for a user should be desired by the user. Thus, the system must intelligently decide what services a user desires at each moment. For example, a user commuting to work will want to know if there is traffic along his route, and in this event he may need an alternate route. In the case where there is not traffic, the user should not be notified. Therefore, the system should know to check for traffic along the user's normal route, and calculate alternate routes in the event of traffic and relay this information to the user, and in the case where there is no traffic the user should not be bothered.

- **A reliable way of integrating all the computing agents into a seamless entity needs to be designed**. The requirement for a transparent system brings about difficult questions and issues that need to be addressed. First, what is an appropriate mobile device for a user? Should this device be a full-size laptop, capable of running applications and providing standard I/O? Or should the device be a thin client, which only provides I/O and outsources heavy computation to resource-abundant machines? As more features are supported by a mobile device, more resources are needed on the device, which implies a physically larger device. Yet, a physically large device violates the transparency requirement. Ideally, these devices would be wearable computers, implying that such a device can be carried as it if were a personal accessory, such as a wallet or wristwatch. Since a small device lacks computational power and battery life, the device must outsource its workload to more capable machines in order to provide the same features of larger devices. Furthermore, the issue of security of operation, when combined with the transparency requirement, raises additional caveats. First, the system needs to be endowed with user authentication procedures that grant a sufficient degree of security while at the same time retaining the invisibility feature of a pervasive application. Secondly, only authorized agents should have the capability of initiating processes for the user; a framework to scan and filter the available services will then have to be integrated into the system.

In this chapter, we will focus on two issues that are crucial for the design of a proactive yet transparent system, in which mobile devices outsource tasks and services to remote machines. First, we will describe techniques for resource management in a pervasive environment, considering a realistic assumption that the proactive issuing of tasks may overwhelm the available resources and may bring about distraction to the user. Second, we will present the issue of authentication and recognition of users and services, again focusing on solutions that minimize the amount of human-machine interaction but still provide the required level of security. We conclude our discussion with the review of a spectrum of current pervasive computing projects.

## 2 Resource Management

As outlined in the introduction, a reasonable strategy to reduce the computational burden of small devices in a pervasive environment is to allow those clients to outsource their workload to larger machines. This outsourcing of work brings about its own issues in the realm of transparency.

The machines that take on the outsourced work are known as surrogates. A surrogate may be a standard workstation, a server-grade machine, or even a cluster of servers. A mobile device may outsource a computationally intensive task to a surrogate through some type of remote procedure call. Thus, the mobile device could act as a thin client, whereby the surrogate runs a user's applications and the mobile device is only used for I/O. For simplicity, this article will group all types of outsourced tasks under the label of *services*. Therefore, the servers and surrogates processing the clients' requests will be known as *service providers*.

A main argument promoting the idea of pervasive computing is that computing power is monetarily cheap, as the most powerful workstations of yesteryear (e.g., late 1990s to 2000) can be bought for a few hundred US dollars. These machines of yesteryear are plentiful, and can easily be set up as surrogates. While this is an easy argument to make about the potential computing power of a pervasive environment, one must also consider the number of users in this environment as well as the demand placed on a surrogate by each user. In an urban or densely populated area, the number of mobile devices requesting services may greatly outnumber the available surrogate service providers, and the service rate of the providers may not meet the needs of the users. Thus, the service providers may form the bottleneck in a pervasive environment. The surrogate machines offering the services can be thought of as a public commodity; therefore it may be difficult to predict their load, and hence, difficult to provision these resources.

While the cost to provision these resources as well as the efficiency of these resources may not be the foremost goals of pervasive computing, there is reason to address these concerns; the bottleneck may degrade the performance to a degree where a user is annoyed by the high latency to perform a task. An element of transparency is therefore lost, as the user becomes aware of the under-provisioned and over-utilized system. Yet, provisioning the proper number of resources may not be possible and the users may have to settle for the best effort provisioning.

Therefore, under best-effort provisioning, we will examine how resources can best be allocated in order to minimize distraction to the user. However, this resource allocation must not require users to micromanage their own resources, as this should be done transparently from the user in a pervasive environment. Yet, this transparent allocation must also provide users with the same level of satisfaction as if the users were managing their own resources. This section will describe techniques to transparently manage resources at the mobile-host level, as well as techniques to alleviate the bottleneck at the service providers.

### 2.1 Distributed Caching

One technique to alleviate the bottleneck and congestion at the surrogate level is to distribute the workload of the surrogates amongst the mobile devices. This is not a novel idea, as its application can be seen in decentralized peer-to-peer (P2P) file sharing systems. In a decentralized P2P system the peers act as both clients and servers, as they request and service queries. Regarding file sharing, a peer sends a request for a particular file, and any peer who has a copy of this file can respond to this request by sending the file to the requesting client. Thus, there is no central server to handle requests, but rather, the clients themselves service each other's requests. Therefore, the workload is distributed amongst the clients instead of concentrated at a single source which would act as a bottleneck.

Regarding a service-oriented pervasive environment, a user sends a request for a common service from his or her mobile device to a remote server which handles the request and provides the mobile client with the response. Common services could be requesting the current weather condition, or requesting the current traffic pattern. In a centralized approach, all mobile clients would send these requests to the same service provider. A provider would process a request, by calculating the most current state of the weather or traffic, and relay this response to the requesting client.

Distributing the load in a service-oriented pervasive environment is subtly different than distributing the load in a P2P file sharing environment. Files in a P2P system are static, and as such, they do not grow stale. The same file can satisfy two requests occurring at two different points in time. On the other hand, the data requested in a service-oriented system is dynamic. Two requests for the weather at different points in time (e.g. a few days apart) cannot be satisfied by the same response. Therefore, one must keep in mind the idea of freshness of the data in a service-oriented environment. For some services (e.g. weather and traffic requests) the service providers are the only ones capable of producing fresh responses to the queries, as these machines are connected to the weather center or news center via a local network or the Internet. A mobile client, on the other hand, only has knowledge of the responses to its own queries as well as those responses to the queries it overhears. It may overhear a query if it is promiscuously listening to the wireless medium, or if the client is being used to forward a query-response pair on the path between the client and server. For other services (e.g. time of day requests) a mobile client may be capable of servicing fresh data.

Thus, while a central server can be eliminated in the P2P file-sharing model since the peers are capable of independently performing a service equivalent to that of the central server, a central server cannot be eliminated in the service-oriented model. The dynamic nature of the data in a service-oriented model suggests that clients will depend of a true service provider to obtain fresh data. A client providing a cached response to a service request is not performing the equivalent service of the true source, since the cache response grows stale over time. Thus, a pervasive environment must incorporate true service providers, and cannot distribute all the workload to the clients themselves. A hybrid model is more appropriate, in which the true service providers are used to service the clients demanding the freshest data, and the clients with cached copies are used to service fellow clients who can tolerate stale data. The following section details the advantages of this hybrid caching model.

### 2.1.1 Advantages of caching in a pervasive environment

The most obvious advantages of employing caching in a pervasive environment are that the bottleneck at the service provider can be alleviated and the response time for service requests can be decreased. The bottleneck is alleviated as requests are dispersed away from the source and directed to the clients. Alleviating the bottleneck lowers the congestion at the source and also lowers the source's workload, which in turn, reduces the response time of a query to the source. More significantly though, response time of a query can be lowered if a mobile client chooses to send its request to another mobile device servicing cached data instead of sending this request to the source. A mobile device serving cached data will be most likely be less congested than the actual source and will also be geographically closer than this source. Thus, response time is decreased due the reduced congestion and a lower round-trip time. However, the most significant advantage of caching is its scalability. As more clients are introduced to the pervasive environment, these clients soon become servers of cached data. Therefore, even though the demand for services will increase due to an increase in the number of clients, the availability of cached data will also increase, and thus, the increase in demand will be met by an increase in supply. One caveat when increasing the number of clients is that as more clients exchange cached data, stale data is propagated to more nodes and the lifetime of this stale data is increased. Stale data propagation will be addressed later in this article, but first, a simple software architecture for implementing a caching system in a pervasive environment will be presented.

### 2.1.2 Software architecture of a caching system

In the protocol presented in [2], mobile devices act as both clients and servers, requesting and responding to service requests. The software architecture of the protocol consists of three entities: Providers, Consumers, and Information Managers. All three entities are housed on a user's mobile device. A Provider process simply offers a service, which it can provide to remote devices or to the local device on which the Provider process is located. A Consumer process requests a service. Lastly, the Information Manager is a process that manages all the Providers and Consumers within a single mobile device, and also communicates with the Information Managers of remote devices. Figure 1 illustrates the communication between these entities.

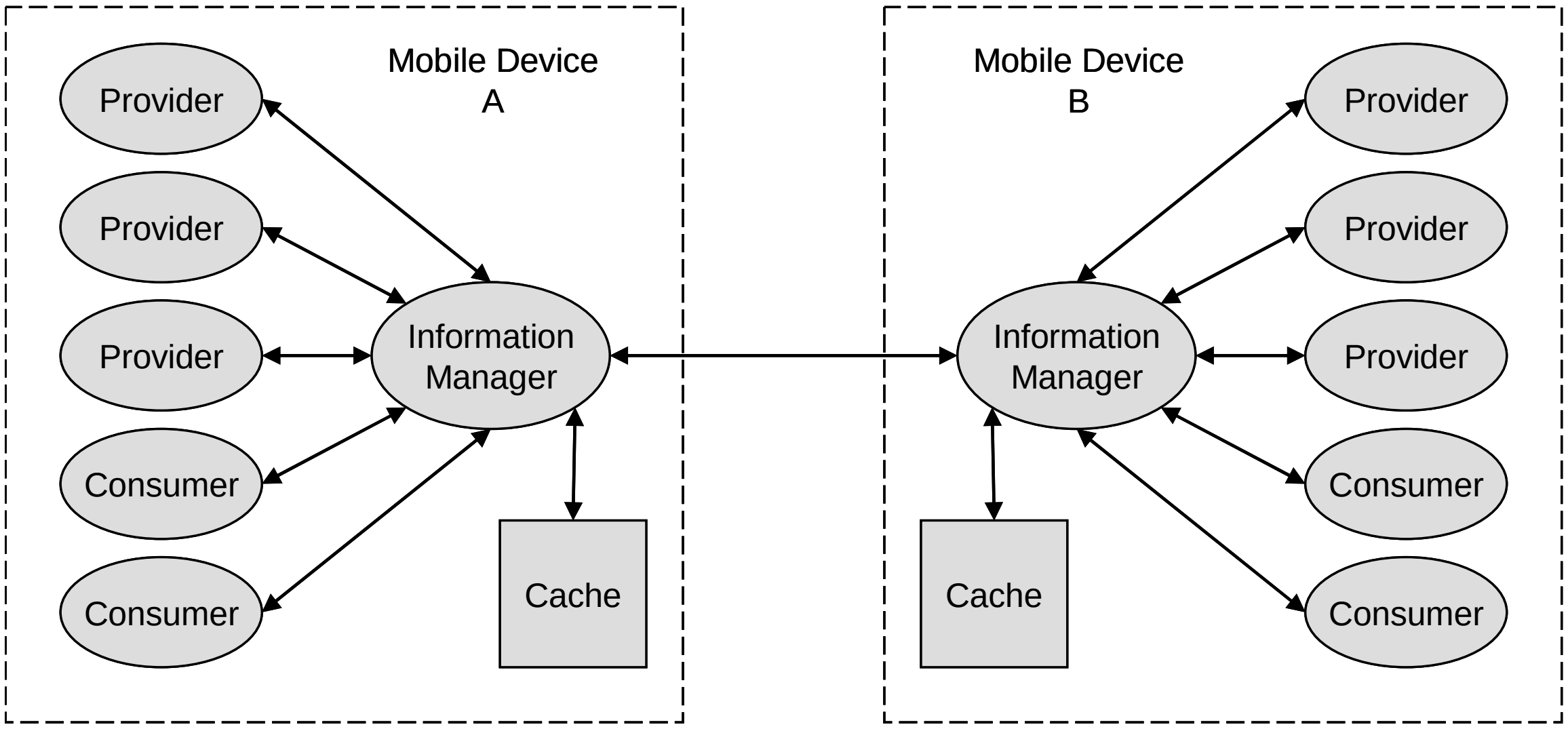

Figure 1 – Communication scheme between Providers, Consumers and Information Managers

In order to provide a service, a Provider will register itself with the local Information Manager. The Information Manager will then advertise this service to all mobile devices within its transmission range (i.e., one-hop neighbors) by broadcasting an advertisement.

When querying for a service, a Consumer will send its query to the local Information Manager of the device. The local Information Manager will first check a cache of queries, and if it has a cached response for the very same query, it will respond with the proper answer. Otherwise, it will check the list of Providers that are registered within this device in an attempt to satisfy this query locally. If no local Provider is found within the device, then the query will be broadcast to one-hop neighbors with the hope that a neighboring device will have a cached response to this query or will have a Provider who can answer the query. In [2], a broadcast query is limited to one-hop as a means to regulate the flooding. As an alternative, one-hop neighbors could subsequently broadcast the query further. However, broadcasting a query aimlessly can easily congest the network, thus it is important to limit the size of the flood. Rather than expanding the flood radius, a host whose query cannot be answered by neighboring hosts may, instead, send its query directly to a known data source, such as a weather center when attempting to obtain the current weather conditions.

### 2.1.3 Propagation of stale data

While having much in common with a decentralized peer-to-peer file sharing protocol, the aforementioned protocol [2] also resembles a proxy cache through its use of cached responses. A proxy cache, or web cache, is located between the true web server and a querying client. It can answer a client's request by serving the client a cached HTML page, instead of forwarding the query to the server. In the protocol offered in [2], each Information Manager can cache queries and responses which it overhears or forwards. If the Information Manager receives a query for which it has cached a response, it will respond with this answer. As with a web cache, the response to this query may be stale. In this situation a tradeoff has been made to serve stale data in an attempt to shift some of the load off the service providers. In [2], simple techniques to deal with stale data are offered. First, cached responses have a timeout period, such that an entry is deleted after some period of time. Similarly, instead of deleting an entry after a timeout, the Information Manager may requery for this service, as a means to update its cache with more current data. Unfortunately, a requery may result in the same stale data being sent by another remote device, and thus the stale data may prolong its existence in the system. These timeout and update schemes are simple, but do not offer the user much control, nor do they guarantee that fresh data will be received after a timeout. Implicit to the pervasive goal, a user's mobile device should be an unobtrusive tool, and as such, it must be elegantly customized and tuned to its user. A user may require more than a simple timeout to govern the updating of data. The user may want a certain level of freshness when accepting cached data, and therefore, the user may wish to know when the cached data was generated by the source. Based on the time data was generated, the user may wish to know how likely it is that the cached data is fresh. The following section offers a different technique to combat stale data, in which the user controls the level of freshness of the cached data that he or she receives.

### 2.1.4 Imposing a quality-of-service metric on cached data objects

The previous sections have discussed the advantages of employing a caching system to alleviate the bottleneck at the service providers, nevertheless, the issue of stale data propagation remains a problem. In general, this issue concerning the spread of stale or fresh data is known as data consistency. Some environments require a strong consistency for cached data. For example, in a multiprocessor environment in which each processor has its own cache, the same data object may reside in multiple caches; yet when any processor reads this object, this read must reflect the latest write to this object. The requirement where any read must reflect the most recent write is known as strong consistency, and thus, stale data will never be read. In a multiprocessor setting, maintaining this condition is important to ensure the correctness of an executed program. Relaxing this condition to a point where the latest read of a data object may not return the value of the latest write yields a weaker consistency. The caching protocol discussed in Section 2.1.2 assumed a weak consistency for data objects, allowing the propagation of stale data. The problem of stale data can be addressed by a technique offered in [7]. While the offered technique was intended for Internet caching, it is just as appropriate in the domain of pervasive computing, as the service-oriented architecture of pervasive computing is similar to the service-oriented architecture of most websites. This technique allows the user to adjust the level of consistency for the requested data objects in order to meet one's personal needs.

It should be noted that maintaining a strict coherency between a data object at the source and all cached copies of the object held at the clients is very time consuming. The classical cache coherency protocols come in two flavors: update and invalidate. These require some service to manage a directory, which stores the list of locations of all cached copies as well as state information describing the current access permissions of the clients. In a pervasive environment the number of clients will likely be too large to manage such a directory. In addition, broadcasting (or multicasting) the invalidate messages, or the larger update messages, would congest the network with great deal of overhead traffic. Moreover, connectivity to the mobile hosts in a pervasive environment is not guaranteed. Thus, the invalidate and update messages are not guaranteed to reach all of the hosts, and therefore, the coherency protocol would only be best-effort. Settling for a controlled weak consistency is therefore an appropriate solution in a pervasive setting.

### 2.1.5 Trading off consistency for response time

When servicing the requests in a pervasive environment where caching is permissible there is a tradeoff between the consistency of the data served and the response time. If a user requires strong consistency, then the request must go to the source serving the data in order to ensure that this user receives the most recent copy of the data (e.g., only the server in the weather center can provide the most recent weather conditions). If we reasonably assume that there is congestion at the main source, or that the mobile client has intermittent connectivity to the source, the client will experience a high response time when querying this source. On the other hand, if a client can tolerate a stale data object, then a cached copy of the object can be received from one of many less-loaded, and possibly physically nearer, clients; thus, reducing the response time. In the technique introduced in [7] the user can decide, for each data object, whether consistency or response time is important.

The decision for consistency or response time is not a binary decision, but rather, there is a continuous domain from which a user can select a point between strongest consistency (high response time) and weakest consistency (low response time) for a particular data object. In [7], this domain is known as a quality-of-service (QoS) domain, and it is illustrated in Figure 2. Values in this domain represent the minimum probability, between zero and one, that the data object received will reflect the latest write to this object. By selecting a value of one, the user is guaranteeing with 100% probability that he or she receives the latest version of a data object. By selecting a value of zero, the user is suggesting that he or she is willing to receive any version of the requested data object, no matter how stale or fresh. By selecting an intermediate value, say 0.3, the user is suggesting that he or she is willing to receive a version of the data object where there is a 30% or greater probability that the received object reflects the most recent version.

Users in a pervasive environment can set a QoS value for each service performed by their mobile device. For example, an investment banker would select a high QoS value for the service updating stock quotes. Only the most current quote is important to the banker. On the other hand, a casual investor may set a low QoS value for the stock quote service. Similarly, a golfer checking the weather at the course may set a low QoS value for the weather service. The golfer may believe that the weather does not change drastically within a day, and therefore a stale cached value would suffice to offer the approximate temperature and weather condition. Yet, a newscaster reporting on a tornado may require the most accurate weather condition, and therefore set a high QoS value for the weather service. Thus, a user must be able to customize this setting for each service.

How can the system determine the probability that a cached object is fresh and reflects the latest write? This probability depends on how often an object is updated and the time of the last update. To implement this, the source for a service would maintain a table of entries, in which an entry is inserted after an update is made to the data object. An entry to the table consists of the time between updates (the time from the last update to the most recent update). An average can be calculated across these entries to acquire the mean time between updates (MTBU) along with its standard deviation ($STDV_{MTBU}$). The source would maintain the MTBU, $STDV_{MTBU}$, and time-of-last-update values and would provide these to a client requesting its service.

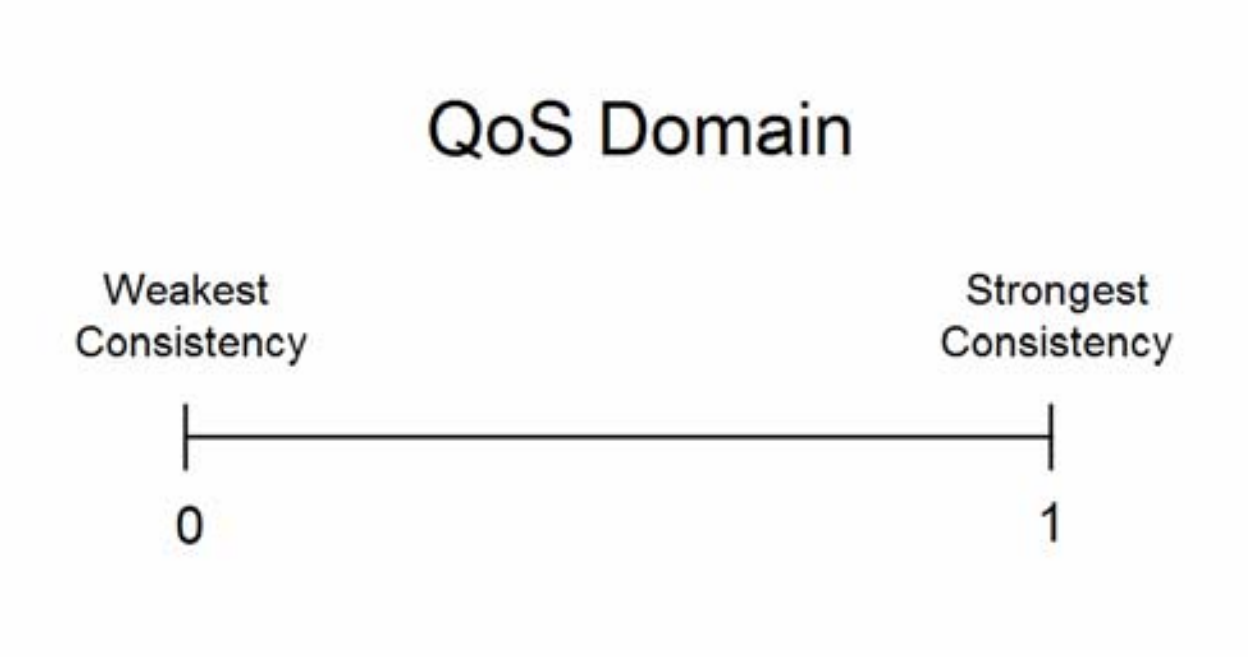

Figure 2 – Quality-of-service domain

A client will use these metadata values to determine the probability that cached object is fresh and has not been modified. When receiving a request for a cached data object, the servicing client can calculate the probability that the data object has been changed since the time the object was cached. This assumes that the time-between-update values follow a normal distribution pattern. The servicing client needs to first calculate how much time has passed since the time-of-last-update. Using this value along with the MTBU and $STDV_{MTBU}$, the servicing client can calculate the area under the normally distributed bell curve for this object (e.g., by integrating over the probability density function or using a simplified statistical z-score table) to determine the probability that the object has been modified ($P_M$). Equation (1) describes the probability density function (PDF) for the time between updates, where the independent variable *t* represents the time since the last update. Figure 3 graphically illustrates this curve. Once the $P_M$ has been determined, the simple calculation in Equation (2) will determine the probability that the object has not been modified ($P_{NM}$).

$$PDF = \frac{1}{\sqrt{2\pi \cdot STDV_{MTBU}^{2}}} \cdot e^{\frac{-(t-MTBU)^2}{2\cdot STDV_{MTBU}^{2}}} \qquad (1)$$

$$P_{NM} = 1 - P_M \qquad (2)$$

Before a request is satisfied, the requesting client should compare the $P_{NM}$ value of a cached object to the QoS value set for this object. If the QoS value is less than or equal to the $P_{NM}$, then the request can be satisfied by the cached data. Otherwise, a different servicing client can be sought, or the true source of the object can be queried. It should be noted that the servicing client and requesting client may be one and the same. Before a client uses an object in its own cache, it should calculate the $P_{NM}$ for the object and compare it to its own QoS value for the object.

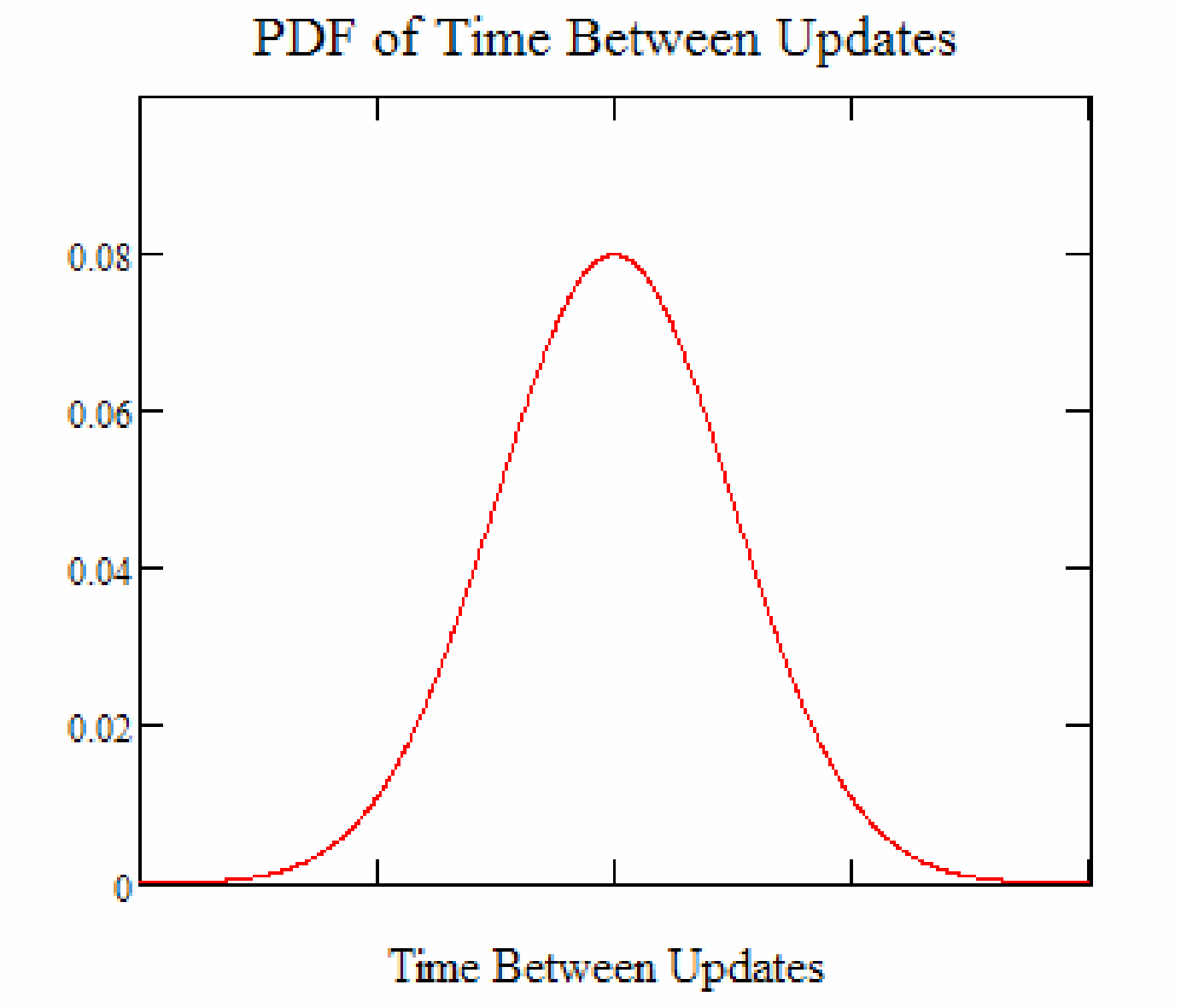

Figure 3 –PDF for the time between updates

### 2.1.6 Alternative QoS metrics

While the proposed method [7] achieves its goal by allowing users to customize the quality-of-service level of their requested data objects, there are alternative metrics for quantifying *quality of service* which may be more appropriate in a pervasive setting. The offered QoS metrics should give the pervasive user control and flexibility, such that the user attains high satisfaction and minimal distraction. The method discussed in Section 2.1.5 uses freshness as the metric for calculating a QoS value. For services reporting data such as stock quotes or breaking news, the freshness metric is well-suited for characterizing quality of service. The changes made to these types of data are important and worth noting. Those individuals requesting these services expect up-to-the-minute data, and thus, the current metric is fitting. However, for a service such as the weather, the freshness metric might not be as fitting. A weather center may update the current temperature, pressure, and wind speed every minute or even every few seconds. Yet, most people do not require an up-to-the-minute report on the weather. Thus, stale data is permissible here. This raises the following question: how should a user set the QoS value for this service? Consider that the user does not mind receiving stale data, as long as the data was generated within the last twelve hours. It would be difficult for the user to set an appropriate QoS value for the object, if QoS values are based on the probability of receiving fresh data. Even once an appropriate probability value is found to satisfy this case, any change in the mean time between updates or standard deviation will cause the range of acceptable objects to change.

Therefore, the time an object was last modified may be a better quality-of-service metric for some data objects; or an even more complicated metric could be devised based on the payload of the data object. Continuing the weather example, an individual typically only cares about significant fluctuations in the temperature. Similar to the mean time between updates, a weather center could record the mean change in temperature and a user could select a QoS value based on this value. On the days where temperature greatly increases or decreases a user would require more current data, and on the days where temperature is stagnant a user could accept older data. Even within a single day, temperature may go through periods of change and periods of consistency; for example, the temperature may greatly increase during the morning, remain stagnant during the afternoon, fall during the evening, and remain stagnant overnight. Thus, even though the weather center may update their temperature reading every minute, a user may only require updates when the temperature significantly changes (i.e., during the morning and evening periods). Therefore, to offer flexibility and customization options to the users in a pervasive environment, multiple QoS metrics should be available.

### 2.1.7 A cache replacement policy

Query resolution based on the quality-of-service of cache contents brings about the issue of a cache replacement policy. In [7] a technique is offered for replacing objects in a client's cache. This policy attempts to keep the most appropriate objects in the client's cache under the assumption that coherence is not maintained between the cached object and the master copy. The issue of coherence changes the definition of an *appropriate object for caching* between the context of a traditional system and that of a pervasive computing environment. In a traditional system, coherence is maintained, and therefore, when describing an appropriate object, spatial and temporal locality are well-suited indicators of appropriateness. Traditionally the least recently used (LRU) policy is employed to take advantage of spatial and temporal locality of objects. In a pervasive environment spatial and temporal locality may still be relevant, yet the opportunity for objects to grow stale should be considered in the replacement policy as well. Objects that are frequently read by mobile clients and infrequently updated by the service provider should be kept in cache, as copies of these objects will remain fresh for a long period of time, and moreover, a client with a cached copy becomes a suitable server for fresh data. On the other hand, those objects that are infrequently read by clients and frequently updated by the provider should be replaced, as these cache copies become stale quickly, and clients with cached copies become poor servers for fresh data.

To create a cache replacement policy based on the previous definition of appropriateness in the pervasive environment, a second set of statistical values needs to be maintained regarding the read history of a data object in order to find the mean time between reads (MTBR). Using the MTBU and MTBR values, a Caching Quality Factor can be calculated, as illustrated by Equation (3). This quality factor is a simple ratio of the MTBU and MTBR. Each cached object will be ranked by this factor; the higher the rank, the more appropriate an object is for caching. Thus, if a new object is to be cached, it must have a higher Caching Quality Factor than the object currently in the cache with the lowest Caching Quality Factor.

$$Caching\ Quality\ Factor = \frac{MTBU}{MTBR} \qquad (3)$$

### 2.1.8 Alternative cache replacement policy

The cache replacement policy outlined in Section 2.1.7 keeps the freshest data objects in cache, and as such, mobile clients become effective servers of cached data to their fellow clients; however, clients in a pervasive environment should not be concerned with serving others, but rather, be concerned with serving themselves. The proposed policy [7] is intended for use in a web cache, where the purpose of the cache is to benefit the global population of clients. A pervasive environment is slightly different, in that the client itself is hosting the cache in order to serve itself first and other clients second. Users in a pervasive environment have the independence to manage their cache however they wish. It can be assumed that users are selfish, and therefore, they want a policy whereby a maximal number of their local requests can be satisfied from their local cache. Assuming that users only accept objects that meet the aforementioned QoS setting, then it can be shown that the proposed cache replacement policy is not optimal for each user. For example, if a user gives a QoS value of zero to an object which he or she frequently reads, the user would not want the object to leave the local cache. Therefore, even if this object has the lowest Caching Quality Factor, it should not be replaced.

A better replacement policy would consider the number of reads one makes to each object (as in the LRU policy), as well as the QoS value for each object, in addition to the MTBU statistics. Using all these factors, a user can calculate if a cached object meets his or her personal QoS setting. If an object meets the QoS value and this object is frequently read by the user then it should be kept, and otherwise it can be replaced. Equation (4) describes an Alternative Caching Quality Factor in the range of [0, 1], where $F_R$ is the read frequency as a percentage cache reads for this object, $P_{NM}$ is probability that the object has not been modified, and QoS is the quality-of-service setting. As long as $P_{NM}$ is greater than QoS, this factor is positive; indicating that the cached object meets the user's QoS requirement. When $P_{NM}$ is less than QoS, this factor is negative; indicating that the cached object does not meet the user's QoS requirement. The greater this factor, the more appropriate an object is for caching. Those objects which do not meet a user's QoS requirement will have a negative caching factor and will be replaced before objects which meet the user's QoS requirement. $F_R$ scales the factor such that an object meeting the QoS requirement is scaled up by its read frequency (i.e., making it more appropriate for caching, indicative of an LRU policy), and an object failing the QoS requirement is scaled down by its read frequency (i.e., making it less appropriate for caching, since it is read often but fails the QoS requirement).

$$Alternative\ Caching\ Quality\ Factor = F_R \cdot \left(P_{NM} - QoS\right) \qquad (4)$$

This alternative cache replacement policy is intended to be user-centric, as users in a pervasive environment are independent, and as such, serving oneself is more important than serving others. Thus, the alternative policy favors objects which meet the individual user's QoS requirements and are read frequently by the user. Consider the aforementioned situation where a user sets a low QoS value for an object. The object will remain appropriate for caching as long as the $P_{NM}$ for that object is greater than its QoS. Thus, even when there is a low probability that an object is fresh, it may be appropriate for one's cache; and the fact that this low-probability object may not be suitable to serve other clients is irrelevant to the selfish user. Additionally, the selfish user wants to retain in cache those objects which he or she frequently reads; and the access patterns of other users are irrelevant. In summation, users are not altruistic and should not base their caching decision on their ability to benefit others, but rather on their ability to benefit themselves.

## 2.2 Broadcasting

As with distributed caching, the simple technique of broadcasting can be used to alleviate the bottleneck and congestion at the service providers. The idea of broadcasting is not unique to pervasive computing, nor is it unique to computing in general. During a broadcast, one speaker delivers information to all the listeners in an area. To apply this idea to a pervasive environment, consider a service provider broadcasting a response of a query to all mobile devices within its wireless range. The mobile devices in range could store a response if it may be of need to the user, or may ignore a particular broadcast. To extend this idea further, consider a service provider broadcasting data that has not been specifically requested, but there is a high probability that this data is in demand by the mobile users. This is the idea behind the broadcasting of analog radio and television. A listener does not request that a radio feed is sent to him or her, but the listener just tunes into the frequency on which the desired station is being broadcast.

The service-oriented environment offered by pervasive computing could benefit from this basic principle of broadcasting. The benefits of broadcasting in a wireless setting have been advocated in [3], [4], and [6]. First, broadcasting scales well to the number of users in a pervasive environment. Consider the example where commuters wish to know the traffic report in their area. Whether there are ten or ten million commuters, the service provider for the traffic report performs the same amount of work when broadcasting this data. Not only does it scale to the number of listeners, but broadcasting can minimize the workload of the service provider, as the service provider only needs to broadcast a data object once in order to transmit it to the listeners, as opposed to responding multiple times to individual queries requesting this identical data object. This same reasoning can be used to argue that broadcasting can minimize the amount of bandwidth consumed. If a broadcast settles more than one potential request for a data object, then it has saved bandwidth by sending this object in one transmission, rather than multiple individual transmissions. Even if a broadcast only settles one potential request, bandwidth has been saved, as an actual request did not have to be sent to the provider. These scenarios show how broadcasting can minimize bandwidth and the workload of the service provider. Yet, if a broadcast data object is not requested by any user, then the broadcast has wasted bandwidth and wasted resources at the service provider. Bandwidth is the targeted resourced in a wireless setting, as wireless bandwidth is much smaller than wired bandwidth, and therefore it must be allocated more efficiently. Another subtle advantage of broadcasting in a wireless setting is that battery-constrained mobile devices can reduce the number of expensive wireless transmissions if they do not need to actively request certain data objects.

### 2.2.1 Published vs. on-demand data objects

In [4], techniques are presented to maximize the efficiency of broadcasting data objects, and therefore minimize wasted resources. The context for these techniques is a general wireless setting, yet the same ideas can be applied to the pervasive environment. Data objects, or services in the pervasive environment, are partitioned into two categories: published and on-demand. A published object is one that is broadcast without request, and an on-demand object is one that is only transmitted upon the request of a user. There must be an elegant balance of objects between these types. If all objects are broadcast all the time, resources will be wasted from the broadcast of those objects that no user desires. Thus, only the most popular objects should be broadcast to minimize wasted resources. In addition, the bandwidth allotted to published objects cannot be so great as to limit the resources for those on-demand objects, thereby increasing the access time for the on-demand objects. The constraints of the analytical model [4] for allocating resources between the two object types are as follows: minimize the number transactions and minimize the access time for a data object. The first constraint implies that a provider should publish as many high-demand data object as possible, in order to eliminate redundant transactions. The second constraint implies that there should be adequate bandwidth allotted to the on-demand requests. Similarly, there should not be too many published objects in order to give each published item a fair share of the published bandwidth. It is apparent that providers benefit from broadcasting via a reduced workload, but users in a pervasive environment also benefit by means of a less congested network due to the elimination of redundant transactions.

In [4], the authors have modeled the access of data objects as a class-based open-form queuing network. Each data object represents a class with its own arrival rate of requests and service rate. Equation (5) shows the expected access time, $t$, for any data object. It is defined in terms of the arrival rates, $\lambda_i$, for each of the $n$ data objects as well as expected access times, $t_{broadcast}$ and $t_{on\text{-}demand}$, for each group of objects. There are $k$ objects in the published group and $n - k$ objects in the on-demand group. Equation (6) is a simplified equation for expected access time for broadcast data objects where $S$ is the size of an object and $B_b$ is the bandwidth allotted to the published, or broadcasted, data objects. It is simplified in that it assumes that all published objects are of the same length, and there are no replicated objects in the broadcast cycle. Thus, the expected access time for published objects is half the cycle time of a broadcast. A more complete equation for the expected access time of published object can be found in [4]. Lastly, the expected access time for on-demand objects, $t_{on\text{-}demand}$, is described by Equation (7). It is defined in terms of the aggregate arrival rate for on-demand objects, $\lambda_d$, and the service rate for on-demand objects, $\mu_d$. The service rate, $\mu_d$, is further defined in terms of the bandwidth allotted to on-demand objects, $B_d$, the size of a data object, $S$, and the size of a request, $R$.

$$t = \sum_{i=1}^{k} \lambda_i \cdot t_{broadcast} + \sum_{i=k+1}^{n} \lambda_i \cdot t_{on-demand} \qquad (5)$$

$$t_{broadcast} \approx \frac{k \cdot S}{2 \cdot B_b} \qquad (6)$$

$$t_{on-demand} = \frac{1}{\mu_d - \lambda_d} \qquad (7)$$

$$\mu_d = \frac{B_d}{S + R} \qquad (8)$$

$$\lambda_d = \sum_{i=k+1}^{n} \lambda_i \qquad (9)$$

With basic calculus, Equation (5) can be optimized in order to minimize expected access time. The optimized equation provides the optimal allocation of bandwidth for on-demand objects, $B_d$, and published objects, $B_b$. Once optimized, the equation can be used to partition a real set of data objects into published and on-demand groups by the following iterative algorithm:

1. All objects are initially categorized as on-demand.
2. The object with the greatest arrival rate of requests (i.e., most demanded) in the set of on-demand objects is moved to the set of published objects.
3. The expected average access time for this configuration of objects is calculated using the optimized equation for expected access time, and the time is compared to some predefined threshold.
4. If the access time is less than the threshold, Steps 2-4 are repeated.
5. Once the access time is greater than the threshold, the algorithm stops and the last configuration to satisfy the threshold is used.

To further save bandwidth, one can batch the responses of on-demand requests [4]. This technique suggests that a service provider should not immediately respond to a single on-demand request with a unicast response to the client. Rather, the provider should wait a small period of time with the hope that one or more identical queries will be received within this time. After this waiting period, a single multicast response can be sent to all clients querying for the same data object. This technique increases the access time for the clients by imposing a waiting time. However, bandwidth is saved whenever multiple requests can be satisfied by a single multicast. Therefore, a tradeoff exists between access time and bandwidth controlled by the waiting time. As the waiting time increases, there is a greater chance for more requests to be received and more bandwidth can be saved in the multicast, at the expense of a greater access time for the pervasive user. This technique is a blend of the on-demand and published realms, as requests are taken in an on-demand fashion, but responses are sent in the broadcast fashion.

### 2.2.2 Broadcast Cells

In [3] and [4], there is insight into implementing a broadcasting system in a real-world wireless setting. For a real-world implementation, geographical areas should be partitioned into cells, whereby each cell has an access point for receiving on-demand requests and broadcasting published objects. Each cell should not necessarily publish the same objects, but rather, each cell should publish the objects which are of greatest demand in that geographical area. For example, in an airport, arrival and departure schedules should be broadcast. In a grocery store, the current sale items can be broadcast. Figure 4 illustrates this idea. To notify users of these published objects, each cell must broadcast a *directory*, which describes a schedule for a time-division multiplexing of data objects. A client can use this schedule to identify the data objects it overhears, as the data objects are broadcast in the order detailed by the directory. When moving between cells, there is no guarantee that the new cell will be broadcasting the same data as the old cell. Even if two cells are broadcasting the same object, the scheduling for this object may be different between the cells; thus for a client to continue to receive a particular service as it moves from one cell to the next, it must read the new directory to determine if the service is broadcast in the new cell and when this service can be retrieved in the broadcast. It should be noted that cells may overlap, and as such, the broadcast of one cell must occur on a different channel(s) than is used by the broadcasts of overlapping cells. The concept of multi-channel broadcasting is described in the following section, and it will become apparent how overlapping cells can share the wireless medium.

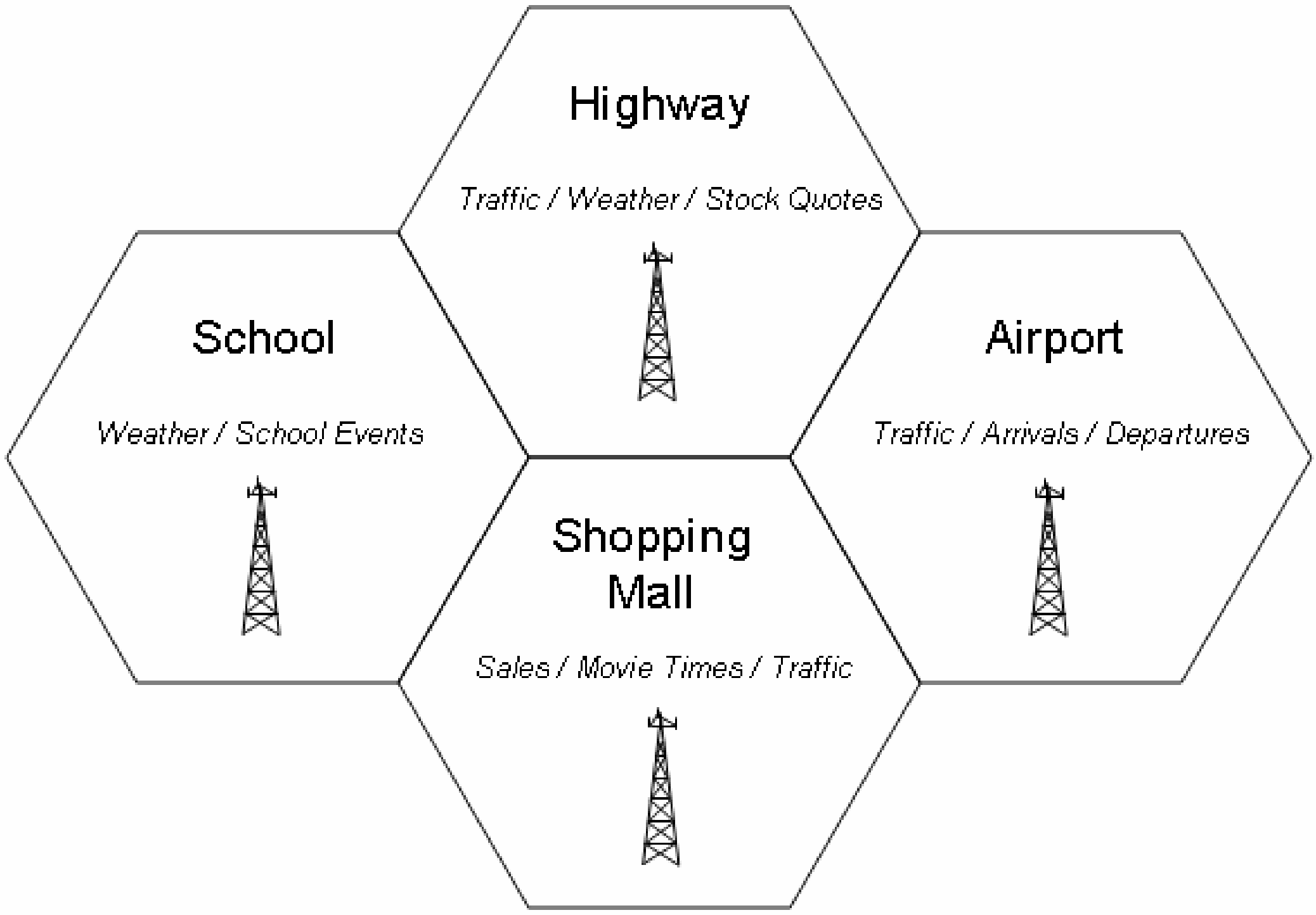

Figure 4 – Partition of a geographical area into broadcast cells (adapted from [4])

Further, the idea of partitioning geographic regions into different broadcast cells is appropriate in many real-world applications, as a person's needs are dependent upon his or her geographical location. One's needs while at work in an urban center are quite different from his or her needs at home in a rural suburb; hence the need for location-aware and location-dependent services. To implement this idea, an administrator of a cell could determine which objects are to be published in a cell, or this set of published objects could be dynamically determined. The aforementioned algorithm for deciding which objects are published and which remain as on-demand is applicable to the situation of dynamically determining the set of published objects in a cell. Under the dynamic allocation, all objects would initially be cast as on-demand, and a fixed number of the most requested objects would move to the published group while maintaining some minimal access time threshold. One caveat of the dynamic allocation is determining if an object of the published group should be removed or replaced after some period of time. Once an object is published, the arrival rate of requests for this object will be low, as only those clients who have not read the directory and are not aware of the published objects would send an on-demand request for a broadcast object. Thus, it is difficult to determine how desirable each published object is to the users when the arrival-rate-of-requests metric cannot be used. Yet, even in the midst of this caveat, it should be obvious that dynamically determining the set of published object is the best solution to meet the needs of users, as these needs vary region by region, day by day, and hour by hour.

### 2.2.3 Balancing response time and power consumption

After decisions have been made as to which data objects should be broadcast in a region and how the airtime should be partitioned into broadcast and on-demand time, there are additional subtle issues which need to be addressed, such as how clients locate requested data objects on a broadcast medium and the ordering by which a client retrieves multiple broadcast data objects. These issues can be discussed in the context of a single-channel or multi-channel broadcast. However, before these issues are discussed, there are two goals which should be kept in mind when considering solutions to these issues.

First, the response time for retrieving data objects should be minimized. This helps maintain the distraction-free environment for the pervasive user. Second, the amount of power consumed by the mobile host should be minimized. A mobile host has a limited battery life, and therefore by reducing power consumption, the lifetime of a mobile host is increased. The mobile host's ability to switch between different modes of operation allows it to save power. In active mode, the host can actively listen to the wireless medium; while in doze mode, the host cannot listen to the medium as its wireless access card is not in use. Therefore, the ideal way for a mobile host to minimize power consumption is for it to only listen to the wireless medium when it is receiving data objects that it desires. During the time when undesirable objects are being broadcast, the mobile host should not listen and should go into doze mode. Since the benefits of pervasive computing are lost when a user's mobile device(s) is powerless, battery conservation is essential. Additionally, the frequent need to recharge one's mobile device creates distraction, which disturbs the seamless integration of technology into one's life. In [6], techniques to address these issues of response time and power consumption are discussed in the context of broadcasting.

### 2.2.4 Indexing

The first issue which will be discussed is known as indexing. Previously, it was mentioned that a directory was to be broadcast along with data objects to describe the ordering of data objects on the broadcast channel. An indexing scheme has same objective: inform the clients as to when a data object will be broadcast. An index for a particular data object may be a hash of certain attributes of the data object, such as a filename or URL. A client desiring a data object will first compute this index by performing a hash of the appropriate attributes. Once computed, this index can be used in a variety of ways in order to locate a data object on the broadcast medium. This section will describe two such schemes for locating data objects: distributed indexing and aggregate indexing. Either indexing scheme will benefit the pervasive user by reducing the amount of power consumed by the one's mobile device, therefore extending the battery life of the device. The point has already been made that one's mobile device is his or her connection to the pervasive computing world, and thus, it is essential that these devices remain powered for a sufficient period of time without recharging.

Under a simple indexing scheme, known as distributed indexing, an object's index is broadcast immediately before its associated data object. To retrieve a desired data object, a client will listen to all the indices on the medium and when it hears the index that matches that of the desired object, the client will retrieve the data object following its index. What are the implications of this in respect to the two aforementioned goals? Without any indexing scheme, the only items that are broadcast on the medium are the data objects themselves. With distributed indexing, index objects are broadcast along with data objects, and therefore, the length of the broadcast has increased. This is illustrated graphically by Figures 5(i) and 5(ii). A longer broadcast length implies a longer response time. The tradeoff for a longer response time is that indexing allows for a reduction in the power consumption of mobile devices. A mobile client only needs to listen to these short indices and the desired data objects; thus when undesired objects are being broadcast the client can save power by switching into doze mode.

As an alternative indexing scheme, the individual indices for each data object can be combined into an aggregate index. The aggregate index can be organized as a serial list of individual indices, describing the order of broadcast data objects much like a directory. The aggregate index can also be structured as a tree, which can be searched faster than a list. The organization of the individual indices in the aggregate index may not be of critical importance and will not be discussed further in this chapter, yet it is important to note that aggregate indexing can further reduce power consumption in respect to the reduction achieved by distributed indexing. Under the distributed indexing scheme, a client has to frequently tune into the medium to hear each individual index, switching in and out of listening mode. Switching modes requires some power. By aggregating the indices, a client only needs to switch into listening mode once in order to hear the entire index; thus, consuming less power.

Next, when should the aggregate index be broadcast? A simple solution is to broadcast this index once at the beginning of each broadcast cycle (see Figure 5(iii)). Under this solution the broadcast length is the same as it is under distributing indexing, yet the average response time is worse. If a client has not yet read the aggregate index, the client will have to wait half of the broadcast length, on average, before reading the index. Then the client will have to wait an average of half the broadcast length, again, to retrieve the desired data object. With distributed indexing, a client only has to wait half the broadcast length for the desired index and data object (since these are always broadcast in sequence); but this decrease in response time is at the expense of an increase in power consumption. Another option is to broadcast the entire aggregate index *m* times throughout the broadcast cycle; this is known as (*1*, *m*) indexing. Figures 5(iv) and 5(v) illustrate this indexing scheme with two different values for *m*. Here, the average waiting time for the index is $L / (2m)$, where $L$ is the broadcast length and *m* is the number of index replicas in the broadcast. The average waiting time for the data object will again be half the broadcast length. This solution increases the broadcast length, but reduces the response time in comparison to the once-per-cycle broadcast of the aggregate index. Power consumption is the same as the once-per-cycle broadcast scheme as well. Thus, the (*1*, *m*) indexing scheme is recommended in [6]. The (*1*, *m*) indexing is suitable for a pervasive environment, since less power is required under this scheme in comparison to distributed indexing scheme. Response time under (*1*, *m*) indexing is only slightly worse than the alternatives.

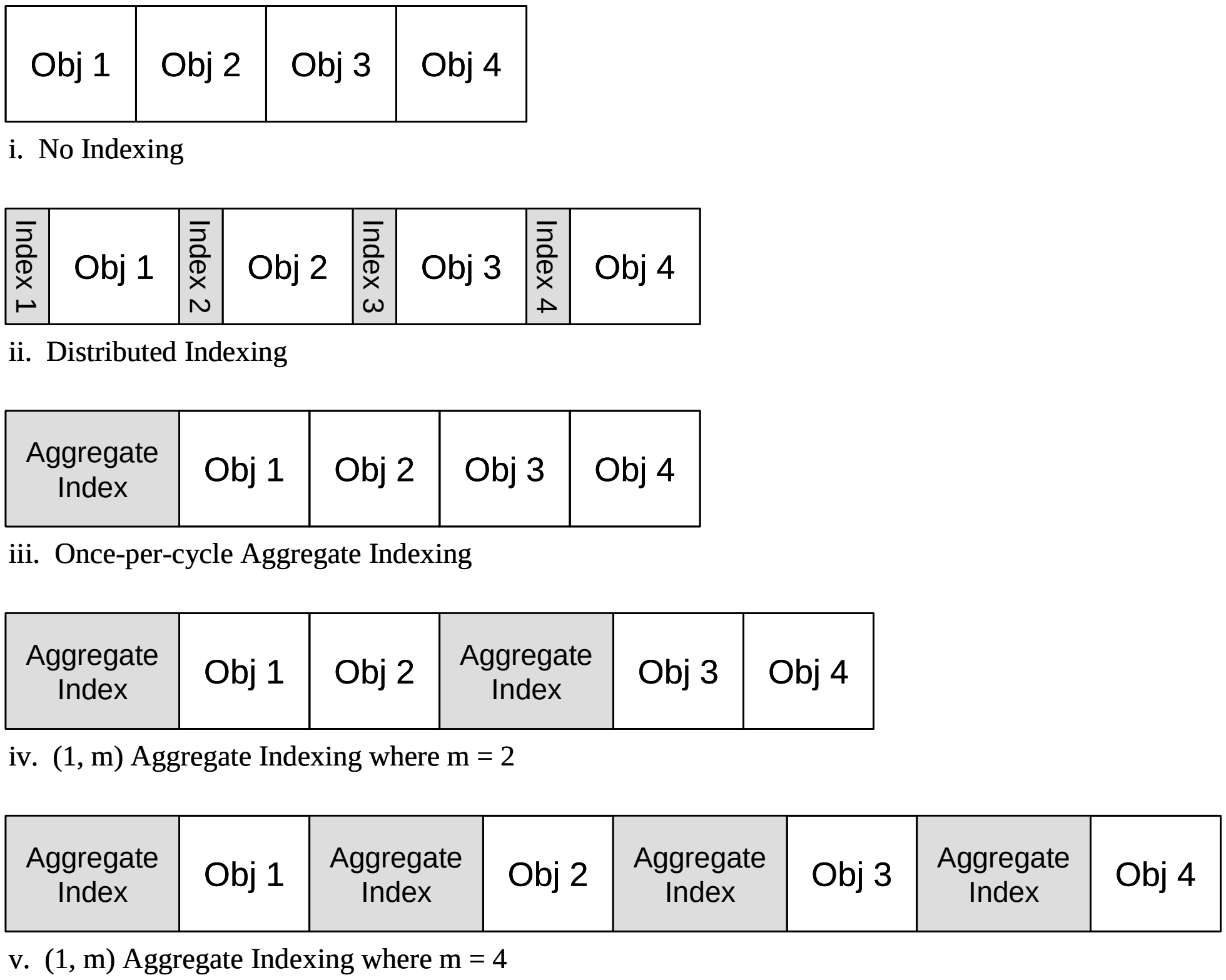

Figure 5 – Graphical representation of indexing Schemes

Thus, when choosing (*1*, *m*) indexing as the appropriate indexing scheme, we are assuming it is more important to the users to extend the lifetime of their mobile device than to experience a slightly lower response time when retrieving broadcasted data. By the nature of the service-oriented architecture in a pervasive environment, response time for broadcasted data does not seem to be critical. The mobile host is expected to predict the user's intent and to retrieve data that it believes the user will want in the near future; thus, data will be ready for the user earlier than if the user were to personally request such data. That is to say, users will not be actively waiting for services to complete, due to the proactive nature of the mobile device in a pervasive setting.

### 2.2.5 Broadcasting over multiple channels

Another issue brought about in [6] which needs to be resolved is whether to broadcast all the data objects on a single channel, or to distribute the data objects across multiple channels. Until this point, a single broadcast channel has been assumed. The tradeoffs of broadcasting over multiple channels will now be explored. The most obvious gain from switching from a single channel to multiple parallel channels is that the length of the broadcast cycle decreases. Assuming $d$ data objects and $c$ channels, each channel must only broadcast $d$ / $c$ objects. Consequently, a shorter broadcast cycle implies a lower response time. Thus, as more channels are introduced, response time decreases. This relationship is true when response time is used to describe the time taken to satisfy a single request. Thus far, response time and power consumption have been analyzed under this single-request scenario. Analyzing the impact of multiple channels in respect to these metrics becomes more complicated in a situation where a client is requesting more than one data object. We must note that in a multiple-channel, multiple-request environment, response time describes the time taken to satisfy all requests.

The main issue when expanding to an environment with multiple channels and multiple requests is the possibility for conflicts. It is assumed that a mobile host can only listen to one channel at a time, and switching channels requires some amount of time and energy. For simplicity, we can assume that each channel is partitioned into the same number of equally sized time slots. Therefore, if two desired data objects are broadcast during the same time slot on two different channels, a client will have to retrieve one object during the first cycle, switch channels, and retrieve the other object during the next broadcast cycle. Even if objects are found in adjacent time slots but on different channels (e.g., $Object_1$ is broadcast on $Channel_A$ in slot $t_i$, and $Object_2$ is broadcast on $Channel_B$ in slot $t_{i+1}$), a client will have to wait for a second broadcast cycle before retrieving the second object. This is due to the fact that channel switching requires some amount of time.

How do channel switching and conflicts affect response time and power consumption? By nature, switching channels consumes power, therefore power consumption increases linearly as the number of channel switches increases. Response time is not so much a function of the number of channel switches, but more a function of the number of conflicts. Yet, it is not as easy to classify the relationship between response time and the number of conflicts. It is obvious when a conflict is introduced in a situation where there are only two requested data objects, response time will increase because a second broadcast cycle will be necessary. Yet, when conflicts are introduced in a setting where more data objects are being requested, additional cycles may not be the consequence. Figure 6 illustrates these situations. Figure 6 (i) shows how an additional broadcast is necessary when there is a conflict between the only two requested data objects. Figure 6(ii) illustrates a more complicated retrieval where multiple data objects are requested, and two passes are necessary for the retrieval. Figure 6(iii) is an extension of the example from Figure 6(ii) with an additional sixth object in the retrieval. The sixth object is in conflict with the first data object retrieved in the first pass. However, this sixth object can be retrieved during the second pass, and therefore, no additional passes were necessary to resolve this conflict. The response time between 6(ii) and 6(iii) remains the same, even though a conflict has been added. In general, conflicts increase response time, but this does not hold in all cases.

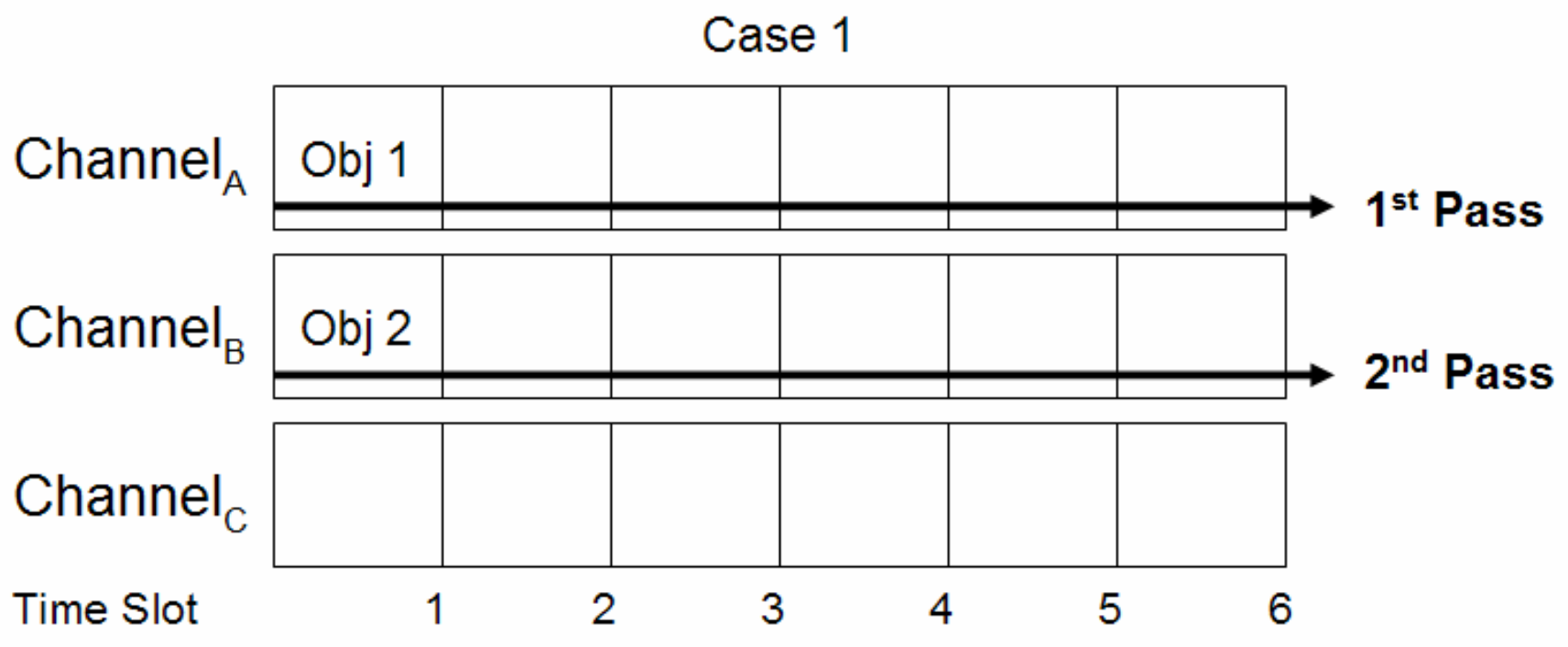

i. The conflict between Objects 1 and 2 results in two passes to retrieve both objects

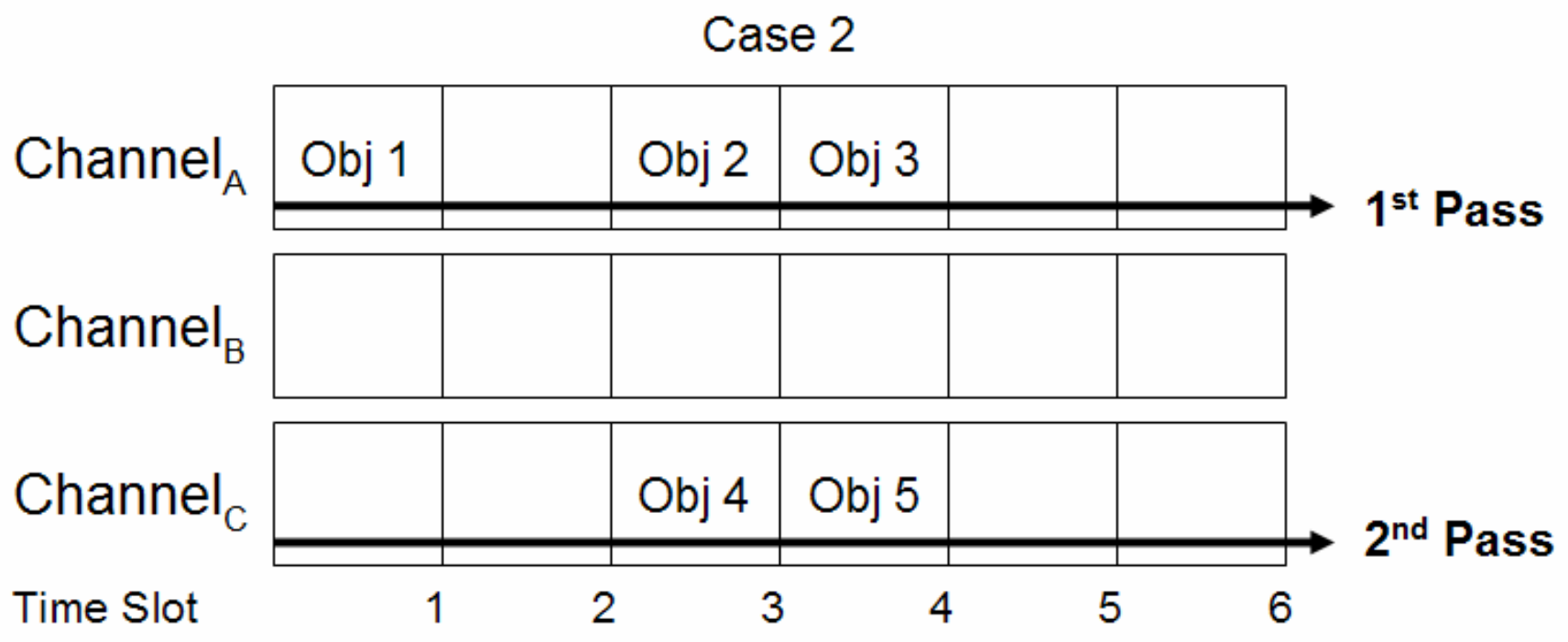

ii. Objects 1 – 5 retrieved in two passes

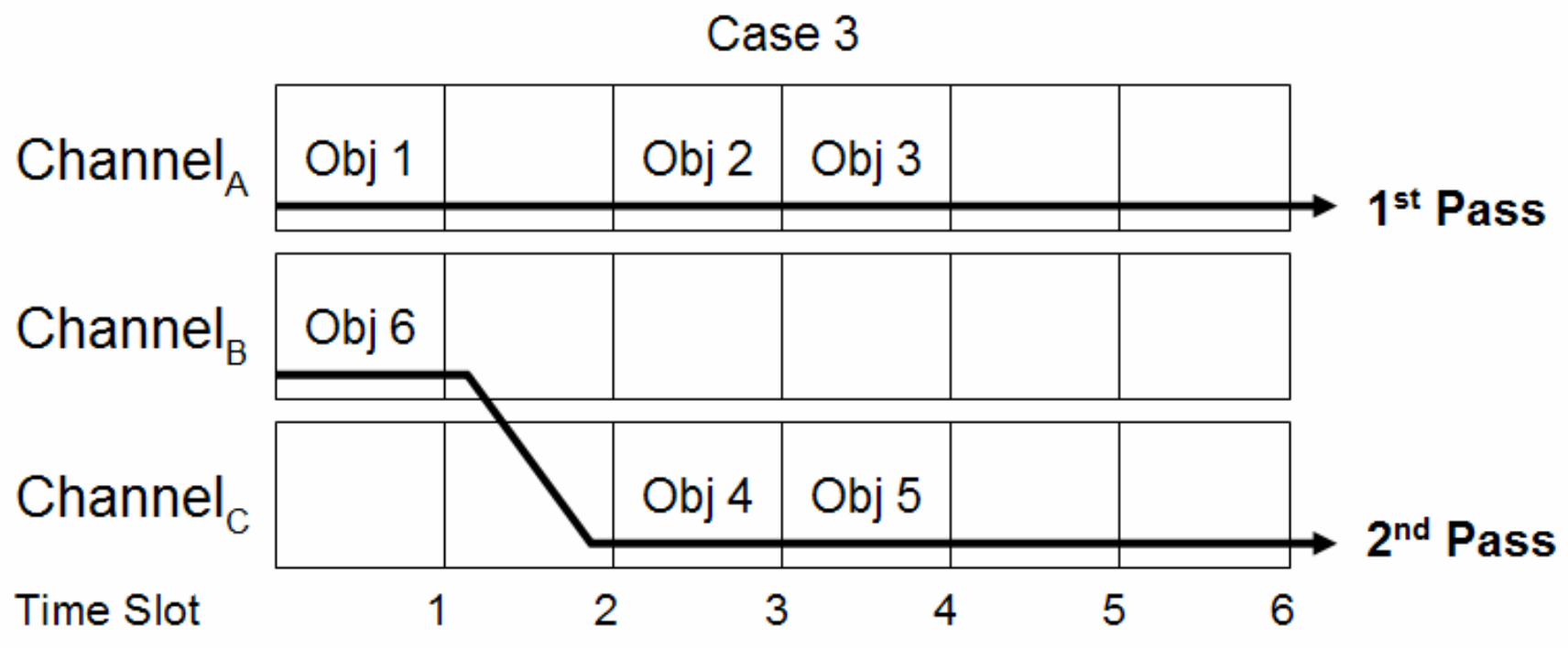

iii. The additional conflict between Objects 1 and 6 does not result in an additional pass to retrieve all objects

Figure 6 – Examples illustrating how conflicts affect response time

It was shown how indexing could reduce power consumption at the expense of a slightly longer response time in the case of a single broadcast channel, yet in the case of multiple broadcast channels, indexing will lower both response time and power consumption. Thus, it is obvious that indexing is beneficial to the pervasive clients under a multi-channel broadcast. First, consider the case where indexing is not used. A client must scan all broadcast channels sequentially to find the desired items. If the broadcast length is $L$ time units and there are $C$ channels, the full scan requires $LC$ time units. When indexing is used, the length of the broadcast cycle will increase, but a full scan may be avoided. Consider the simple case where only one object is requested. A client may read an aggregate index from $Channel_A$ and skip to $Channel_D$ to read the desired object in the same broadcast cycle. Thus, scans through the intermediate channels were avoided and response time is lower than it would be under a situation where indexing were not used.

Therefore, for implementation in a pervasive environment, it is beneficial to implement multiple broadcast channels along with an indexing scheme in order to lower the response time for service requests as well as reduce the power consumption of the mobile device. However, the number of channels implemented in a real-word pervasive environment depends on the resource availability. For a wireless medium, the number of channels depends on the number of available radio frequencies. Analogously, the bandwidth of a fiber network depends on the number of fiber cables laid. It is obviously advantageous to increase bandwidth, yet the amount of fiber laid depends on financial resources. This relates back to the motivating factor for this section on resource management. Even if computing power and technological resources are cheap, they are not endless; hence, resource management techniques are necessary. To the point of broadcast channels, a maximal number of channels should be implemented in respect to financial and physical constraints. Once the number of channels has been determined, a (*1*, *m*) indexing scheme can be implemented on each channel to describe the broadcast objects. As an alternative, one or more broadcast channels can be solely dedicated to broadcasting indices, while the other channels broadcast only data. Under this solution, the waiting time for an index is reduced, and hence, response time for a data object is reduced at the expense of a dedicated index channel(s).

### 2.2.6 Retrieval algorithms

This section will describe three algorithms for retrieving data objects on a multiple-channel broadcast medium. While the details of these algorithms can be found in [6], this section will only present the basic idea of each algorithm in order to find a fitting algorithm for a pervasive environment. With each algorithm it is assumed that the client has read the aggregate index and knows the broadcast location of each desired data object. Knowing these locations, the retrieval algorithms attempt to find an optimal order by which to retrieve the objects.

The first, and simplest, algorithm is a *Row Scan*, where the term *row* is synonymous with *channel*. Here, a client tunes into each channel, on which desirable objects reside, for one pass of the broadcast cycle. During the pass, the client will retrieve each desired object that is broadcast on the current channel. After one pass, the client will then switch to the next channel on which desirable objects reside. Figure 7(i) illustrates this algorithm. The Row Scan provides the client with a minimal number of channel switches. It also requires minimal computation to determine the ordering; a client simply uses the index to determine which channels are broadcasting desirable objects, and sequentially cycles between these channels.

Secondly, there is a greedy algorithm known as *Next Object Access*. Under this algorithm, a client begins by retrieving the first available object with respect to time. It then continues to select the next earliest object which can be retrieved, no matter which channel that next object resides. Figure 7(ii) illustrates this algorithm. As a greedy heuristic, it does not guarantee optimal performance. This algorithm will usually require more channel switches than a Row Scan, and at best, it will require as many channel switches as the Row Scan. Thus, power consumption is greater under this algorithm. Simulation results from [6], shown in Figure 8, find that this algorithm provides a lower response time than Row Scan when the number of requested object is low; and yields a higher response time than Row Scan as the number of requested object increases. Figure 8 plots response time as the number of broadcast cycles needed to retrieve all requested objects.

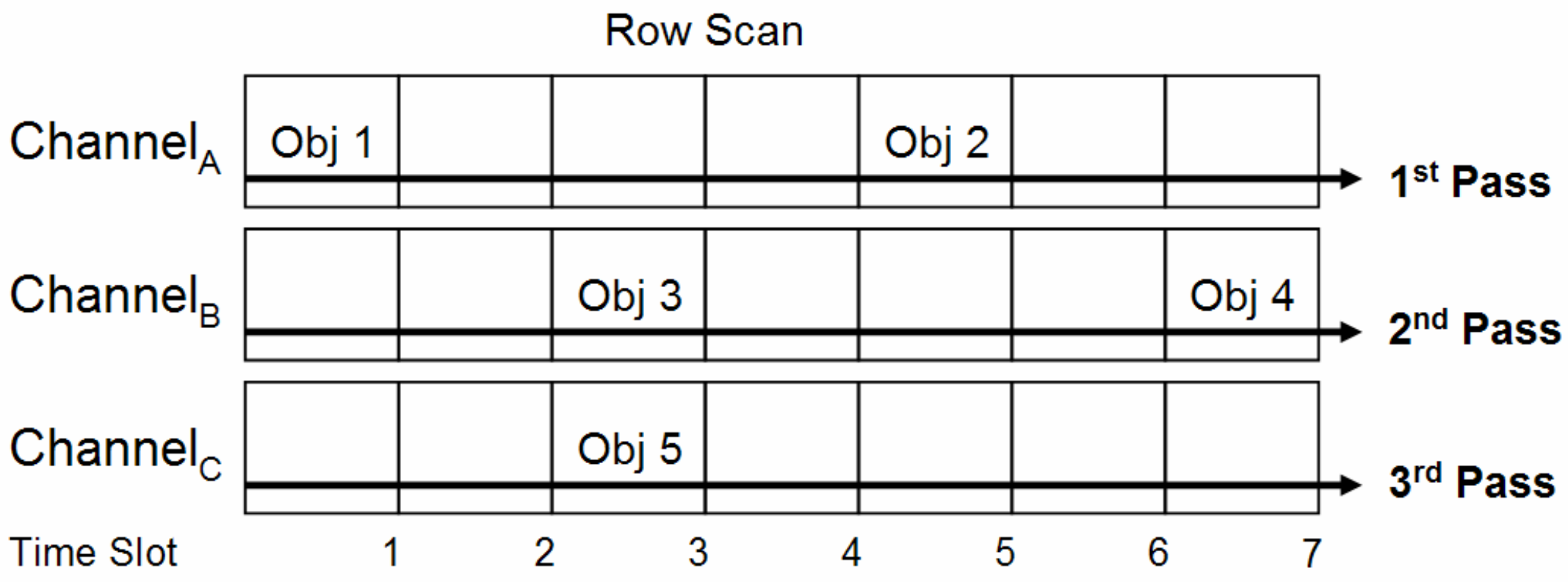

i. Objects 1 – 5 are retrieved in three passes, and two channel switches

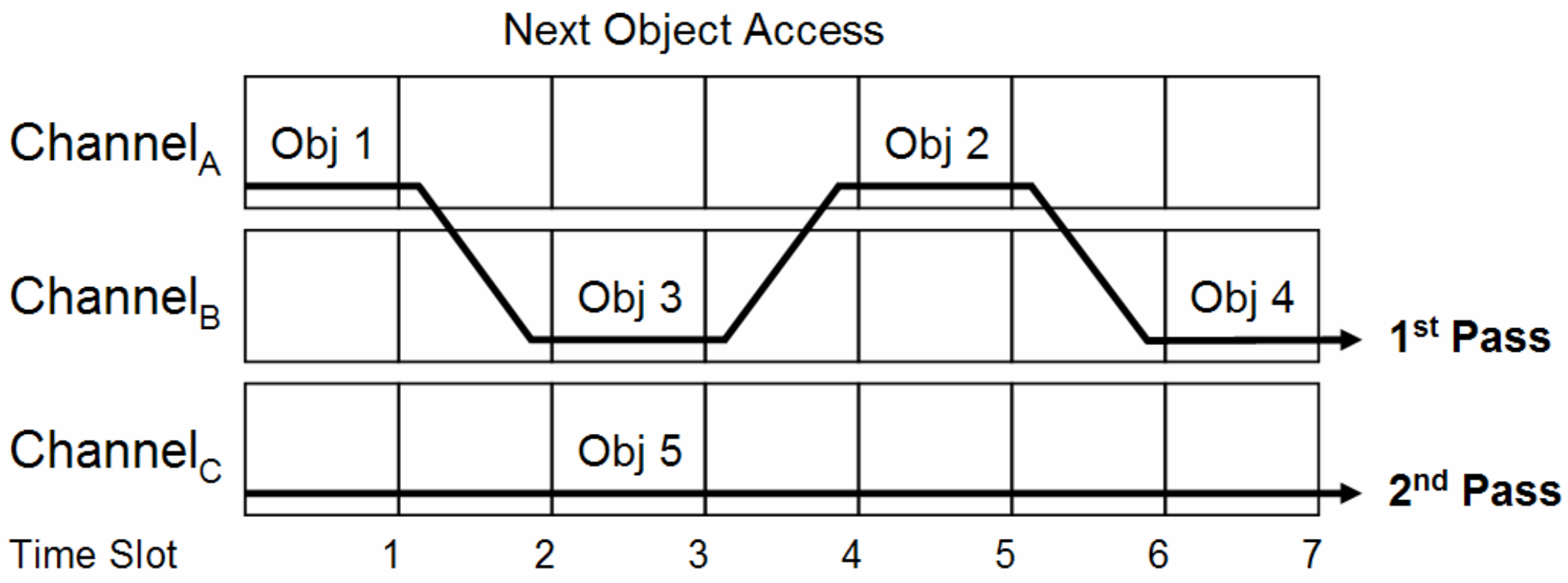

ii. Objects 1 – 5 are retrieved in two passes, and four channel switches

Figure 7 – Row Scan and Next Object Access retrieval algorithms

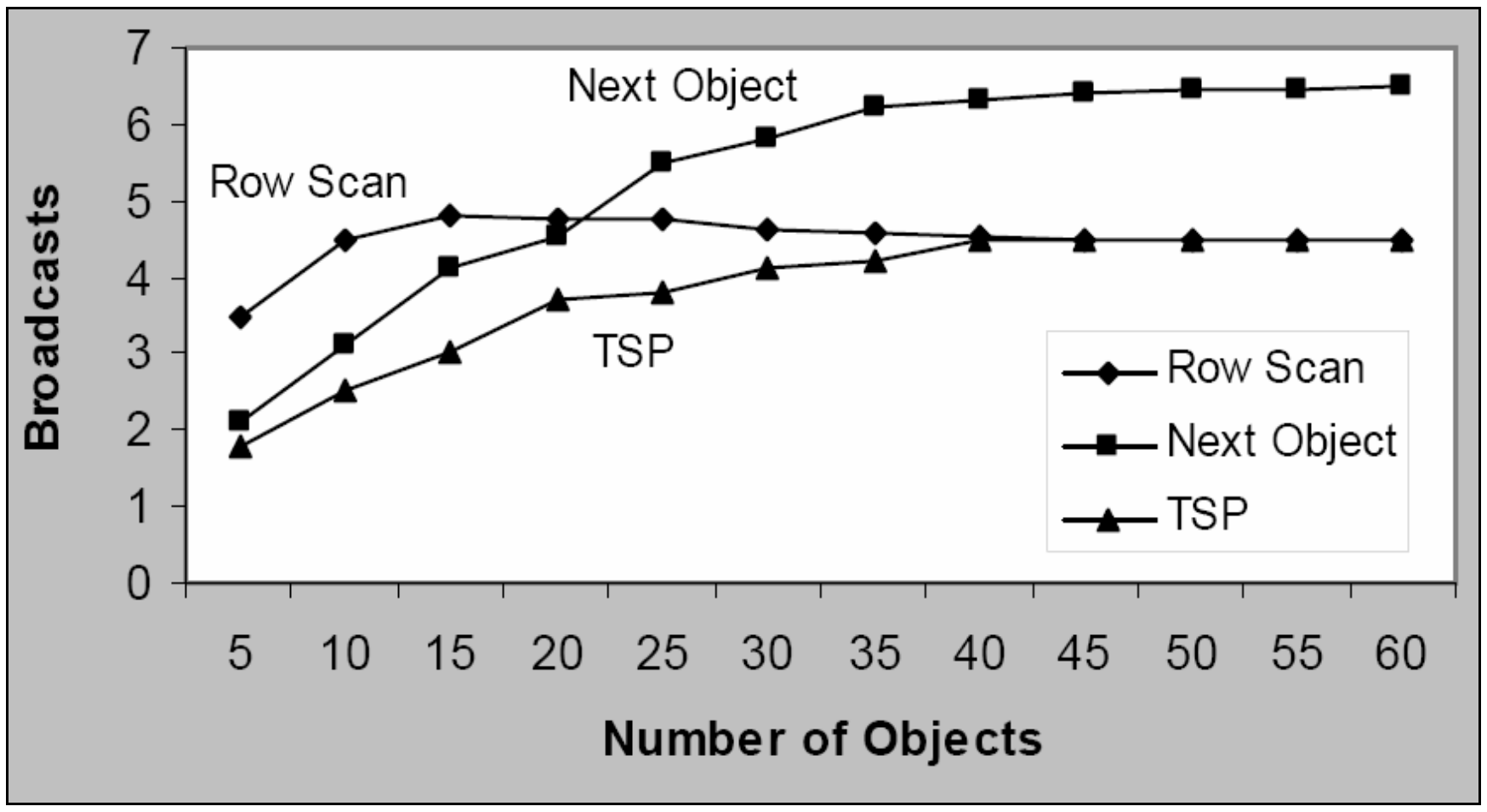

Figure 8 – Simulation results from [6] illustrating how response time varies with the number of requested objects

Lastly, finding the optimal retrieval order of objects in a broadcast is similar to finding the optimal order of cities to visit in the classical traveling salesman problem (TSP). The goal of TSP solutions is to avoid an exhaustive search through all possible orderings of data objects. Simulation results, shown in Figure 8, find that TSP heuristics yield a lower response time than both the Row Scan and Next Object Access algorithms. Further, the number of channel switches under the TSP algorithm will be greater than or equal to the number of channel switches of the Row Scan algorithm, therefore, power consumption will not be improved from that of the Row Scan.

Which retrieval algorithm is most appropriate for a client in a pervasive environment? Because users have different needs, the easy solution is to push this decision to the individual user. A user could select some point between minimal response time and minimal power consumption, and the most appropriate algorithm can be chosen. This would be similar to the aforementioned technique for choosing QoS levels for cached data objects, presented in Section 2.1.5. However, regarding the retrieval of broadcast data, the users in a pervasive environment would most likely value power consumption and computational cost over response time. As argued in Section 2.2.4, since a mobile device is to predict a user's intent and request services in advance, the response time of these services is not crucial. Reducing power consumption, on the other hand, is important to extend the battery life of a mobile device. Additionally, computational cost is important to the pervasive user. Given that a mobile device independently performs services for its user, it is safe to assume that many services and decision-making processes will be competing for computational time on a user's mobile device. Thus, the choice for a power-conserving, computationally-simple algorithm which sacrifices response time may be appropriate. The Row Scan algorithm fits this description, for it has the lowest computational cost and power consumption, and yields a response time that is close to that attained by the TSP heuristics, especially when the number of requested objects is high.

## 2.3 Adaptive Fidelity

Another resource management technique which can help create a distraction-free user environment is fidelity adaptation. The term fidelity is used to describe the quality of an offered service; thus, fidelity adaptation implies a dynamic adjustment in the quality of a service in order to meet the current resource constraints. Lower fidelities require fewer resources, while higher fidelities require more resources. Examples of fidelity parameters include frame rate and resolution for a streaming video service. Continuing this example, a streaming video player may lower the frame rate or resolution if bandwidth is decreased due to a congested network. If the bandwidth increases, the frame rate and resolution can be increased. For traditional services, users can control these fidelity parameters to meet their needs. A user may request low resolution video stream if his or her bandwidth is low, or may request a web page without images in the same scenario to reduce the latency. In a pervasive environment a user should not be bothered by manually setting fidelity parameters each time resource levels change. Micromanagement of fidelities takes away from the transparent user experience, in which technology is seamlessly incorporated in the everyday activities of the user. Thus, the system must make dynamic decisions about fidelity without user input. There are two questions which the system must answer when making decisions on fidelity adaptation: How do different settings of fidelity parameters affect resource consumption? What are the user's preferences in terms of which fidelity factors to adjust in a given situation? Section 2.3.1 will address the first question by presenting a technique to determine resource consumption as a function of fidelity parameters. Section 2.3.2 will address the second question by presenting a technique that determines the level of user satisfaction (i.e., utility) provided by a configuration of fidelity parameters.

### 2.3.1 Modeling resource consumption as a function of fidelity

The authors of [5] have designed an empirical method for estimating resource consumption as a function of fidelity. This method captures and logs resource usage levels as well as the associated fidelities during the run of an application. Based on these historical values, resource consumption can be modeled as a function of fidelity at the expense of computation and storage overhead. As an alternative to this technique, an analytical model can be used, whereby the application developers provide a function for resource consumption in terms of the fidelity factors that their application offers. Yet, it would be difficult for developers to construct a function that is generic enough to apply to a variety of machines with different hardware configurations. In order to tailor such a generic function to any hardware configuration, the function would have to incorporate a sufficient number of input parameters that can be used to describe the architecture and organization of any machine (e.g., processor speed and memory size are two such input parameters). Unfortunately, systems are defined by more parameters than simply clock rate and memory size. Other important parameters include memory organization, processor organization, processor architecture, cache hit rate, and many more. These parameters all affect how resources are consumed, but are difficult to incorporate into a function. The empirical model in [5], however, does not need to consider any of these parameters, as it only uses historical statistics as its basis for estimating resource consumption. Thus, its simplicity makes it a better choice than the complicated analytical approach. The empirical approach can be applied to any application on any machine.

The empirical method implemented in [5] is broken into three phases: logging, learning, and an online phase. During the logging phase, a service is run at different fidelity levels while the resource consumption is monitored by hardware and software monitors. The learning phase uses the logged results to build an estimation function for resource consumption in terms of multiple fidelity factors. For an inexpensive computation, a linear regression is used to map the fidelity parameters (e.g., frame rate and resolution) to the consumption level of a single resource (e.g., computation time or bandwidth). At the end of the learning phase, each resource is estimated by its own function. Initially, the logging and learning phases are performed offline. Thus, when a user first uses a service, the logging and learning phases will have been performed, such that during the online phase, the fidelity parameters of the service can be tuned to meet the resource constraints. In order to tunes these parameters, the current resource constraints (e.g., available CPU and bandwidth) become the input parameters to the functions, and the maximum fidelity levels that satisfy these constraints are returned.

Equations (10) through (13) show an example of the set of functions determined by the learning phase. In this example, the consumption of four resources is modeled by four functions with respect to three fidelity parameters. During the online phase, the availability of each resource will be monitored and will become the input to these functions. The functions will be used to determine the appropriate levels of fidelity to satisfy these resource constraints. Consider the case where $Resource_1$ represents bandwidth, and there is 10Mbps of available bandwidth. Equation (10) will then be fixed to 10Mbps, and solved to determine the configurations of the three parameters (i.e., $Param_1$, $Param_2$, $Param_3$) that can satisfy the constraint of 10Mbps. Consider that these parameters are frame rate, resolution, and audio quality. There will be multiple configurations of frame rates, resolutions, and audio qualities that satisfy the bandwidth of 10Mbps. Which configuration is most appropriate? As available bandwidth decreases, should frame rate, resolution, or audio quality degrade? In [5] it is assumed that there exists a utility function that details a user's preference between fidelity parameters. Hence, the utility function may show that a user prefers a drop in frame rate to a drop in resolution as bandwidth is reduced. The following section will take a deeper look into techniques to estimate the utility functions of a user.

$$Resource_1 = f_1(Param_1, Param_2, Param_3) \quad (10)$$
$$Resource_2 = f_2(Param_1, Param_2, Param_3) \quad (11)$$
$$Resource_3 = f_3(Param_1, Param_2, Param_3) \quad (12)$$
$$Resource_4 = f_4(Param_1, Param_2, Param_3) \quad (13)$$

### 2.3.2 Modeling utility as a function of fidelity

As part of Project Aura [9] at Carnegie Mellon University, an analytical model has been developed to determine the configuration of fidelity parameters that maximizes a user's utility in the presence of resource constraints [8]. The analytical model efficiently finds an optimal configuration of fidelity parameters for a service by considering a set of service providers, a set of fidelity configurations for each provider, and the utility provided to the user by each configuration. For example, fidelity configurations for a video player would be the pairs of frame rates and resolutions that satisfy a given bandwidth. It is assumed that given a fixed resource level, it is possible to find all the configurations of fidelity parameters which satisfy the resource constraint. The functions constructed by the empirical technique presented in Section 2.3.1 can provide such configurations. Once this set of configurations is known, the configuration which provides the user with the greatest utility should be chosen.

The proposed model calculates a single utility value for each configuration of fidelity parameters. Utility assumes a value in the range of [0, 1], where zero implies that a user is completely unsatisfied, while one implies that the user is completely satisfied. The simplest way to map utility values onto fidelity configurations is to have the user choose a utility value for every possible configuration. However, the number of possible configurations can be undoubtedly large. For example, the set of configurations for a streaming video player is the Cartesian product of the frame rate and resolution offerings. Consider that the set of offered frame rates includes 20, 30, and 40 frames per second (FPS), and the set of resolutions includes the qualitative values *high* and *low*. The Cartesian product of these two fidelity parameters produces the complete configuration domain shown in Table 1. This domain grows quickly when either the number of fidelity parameters is increased (e.g., adding a third fidelity parameter of audio quality) or the set of offered values for a particular fidelity parameter is increased (e.g., adding 50 FPS to the set of offered frame rates). To avoid this large configuration domain and its rapid growth, each fidelity parameter can be considered independent from one another. Therefore, instead of mapping a utility value to each configuration in complete configuration domain, a utility value only needs to be mapped to each offering in the individual fidelity-parameter domains. In our continued example, instead of mapping utility values to each of the six possible configurations in the complete configuration domain, utility values only need to be mapped to each of the three frame rate offering, and each of the two resolutions.

For discrete fidelity-parameter domains, such as the ones presented in the Table 1, a mapping table suffices to capture the mapping between utility and fidelity. These mapping tables would need to be manually set by the user in a pervasive environment. These could be set once in an initial offline setup procedure. The user would not be distracted by such configuration decisions afterward the initial setup. For continuous domains, such as the volume of an audio track, a mapping table cannot be used since there are an infinite number of elements in these continuous domains. Instead, a sigmoid function can be constructed to capture the relationship between utility and fidelity over a continuous domain. Figure 9 depicts a utility function in the form of a sigmoid function. The function asymptotically approaches a lower limit and an upper limit. By knowing the range and the knees of a sigmoid function, a continuous function can be interpolated. Therefore, storing a utility function for a continuous domain parameter only requires storing the two knee values, as the upper and lower limits are known to be one and zero, respectively. Upon setup, a user needs to manually determine the knees. These can be prompted to the user by asking him or her for the domain value which is insufficient for the service (lower knee) and a domain value which is good enough for the service (upper knee).

Table 1 – Discrete fidelity parameter domains

| | |
|---|---|
| Frame Rate Domain | {20, 30, 40} |
| Resolution Domain | {*high*, *low*} |
| Complete Configuration Domain | {(20, *low*), (30, *low*), (40, *low*), (20, *high*), (30, *high*),(40, *high*)} |

Equation (14) is the basic formula for determining the maximum utility over a set of service providers and fidelity configurations within each provider. The inner most term, $F_p^{w_p}(c_p)$, is the aforementioned utility function for an independent fidelity parameter. This function is either given by a mapping table or a sigmoid function. The power term, $w_p$, is a weight denoting how important a fidelity parameter is to the user. This is also a manually set parameter in the range of [0, 1], where zero is the highest weight and one is lowest. This weighted utility value is calculated for each fidelity parameter in a configuration and the results are multiplied together; the resulting product will be between [0, 1]. This result denotes a user's overall utility from a single configuration offered by a service provider. A consequence of multiplying the individual utility values together to form the overall utility is that if the utility received by a single parameter in the configuration is zero (i.e., completely insufficient), then the overall utility will be zero and will indicate a completely insufficient configuration. Lastly, this product is then multiplied by the value, $F_s$, indicating a user's preference for the service provider. Again, this value is manually set by the user in the range of [0, 1], where zero denotes that the provider is insufficient and one denotes that the provider is completely sufficient. A user may prefer one service provider over another due to some qualitative features of the providers which are not captured by the utility function. For example, Emacs and Microsoft Word are both word processors, but a user may prefer Microsoft Word over Emacs because of the formatting options offered by Microsoft Word. The value, $F_s$, allows a user to express these preferences.

$$\text{Maximum Utility} = \max\left[ F_s \cdot \left( \prod_{p \in QoS\ Param(s)} F_p^{w_p}(c_p) \right) \right] \qquad (14)$$

When finding the maximum utility value, the work of an exhaustive search through the configuration space can be reduced by implementing an elegant stop condition. First, service providers are to be visited in descending order of their $F_s$ values. A utility value for each configuration is calculated and the greatest value is saved as the current maximum. Before repeating this process for the next service provider, the $F_s$ value of the next service provider is compared to the current maximum utility value. If the next $F_s$ value is less than the current maximum, the algorithm can be stopped, as the current maximum will be the global maximum. This is true because the aggregate product term is upper-bounded by one, and therefore, the overall utility is upper-bounded by $F_s$ for a service provider. Therefore, a service provider does not need to be considered if its $F_s$ value is below the current maximum. By visiting the providers in descending order, it is safe to stop the algorithm once this condition is met.

The proposed algorithm achieves its goal, but it has its limitations. The analytical model offers a solution to find the configuration of fidelity parameters that offers the highest utility. It is distraction-free during the online phase, but requires the user to set: (i) utility values for the elements in the individual fidelity domains; (ii) weights for each fidelity parameter; and (iii) preferences for each service provider. It makes the assumption that by considering each fidelity parameter independently, the optimal global configuration will be found. Lastly, its early termination condition can only be used if a user has different preferences towards service providers. When using this condition, the algorithm will only show a great improvement over an exhaustive search if there is significant deviation in the preferences towards service providers. When dealing with homogeneous service providers, those for which a user has equal preference, an exhaustive search must be performed. Therefore, while the analytical approach may reduce the storage by assuming independence of fidelity parameters, it has not been proven that fidelity parameters should be interpreted independently. Further, the search time can only be reduced when there is enough deviation between a user's preferences towards service providers.

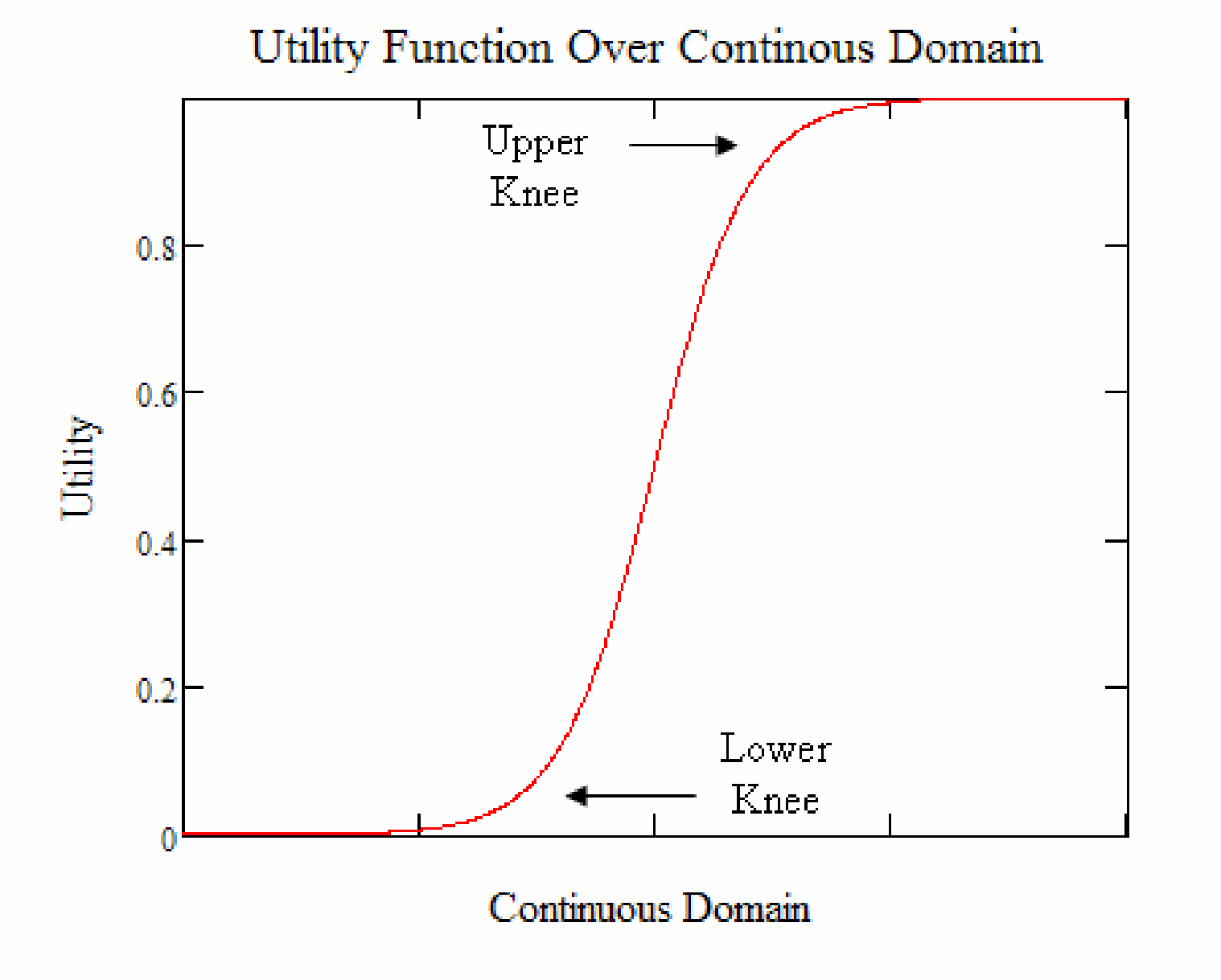

Figure 9 – Utility function over a continuous domain

## 2.4 Conclusions

This section covered three solutions to the resource management problem in pervasive computing: distributed caching, broadcasting, and adaptive fidelity. One motivating factor for these solutions comes from the fact that pervasive computing is intended to be service-oriented, and thus, there is foreseeable congestion at the service providers. Therefore, to prevent congestion at the service providers, the solutions must be scalable with respect to the number of clients which are sending requests to these providers. The techniques of distributed caching and broadcasting were offered to alleviate this congestion in a scalable manner. A second motivating factor for these solutions comes from the fact that many of the clients in a pervasive environment will be mobile devices, whose resource capacity is low. Resources such as battery power, wireless bandwidth, CPU time, and memory must be used efficiently in order for the mobile device to provide the user with a distraction-free experience. Realizing that the independent actions (i.e., service requests) made by a mobile device may place a high demand on system resources, a technique was offered to adaptively adjust the fidelity of the services in a manner that provides the user with the most utility. Resource conservation, specifically regarding power consumption, on the mobile device was also considered when choosing the appropriate implementation for a broadcasting system in a pervasive environment.

Distributed caching shows its scalability in the fact that clients become service providers for cached data. As the number of clients increases, the number of service providers increases as well. However, the caveat is that cached data becomes stale over time. To combat stale data, a technique was presented to allow users to select some quality-of-service level for the data objects they receive. For some data object, a user may sacrifice response time to obtain the most recent version of the object, or may relax the constraint for fresh data, thereby decreasing the response time. The mean time between updates was the key metric in defining quality of service.

Broadcasting proves its scalability, in that, as the number of clients receiving broadcast data increases, the resources required to broadcast this data remains the same, since the service provider does not process requests for this data, nor does the provider need to transmit more data onto the network as the number of clients increases. An algorithm was presented to partition the bandwidth between broadcast and on-demand data objects, in order to minimize the access time of requests. Further, the idea of indexing was presented as a solution to minimize power consumption at the clients, with the only tradeoff being an increased response time under a single-channel allocation. As the number of channels increases, response time decreases, but there becomes the necessity to implement a proper retrieval algorithm to acquire all desired data objects in the face of conflicts. In a pervasive environment, the choice for a retrieval algorithm as well as for an indexing scheme depends on the tradeoff between power consumption and response time. By slightly sacrificing the optimal response time, indexing can be implemented to greatly decrease power consumption, and the Row Scan retrieval can be employed to offer the minimal number of channel switches. Since one's mobile device is intended to issue requests without user interaction, response time is not crucial, for the user will not be actively waiting for each issued request. The effect of power consumption on the battery life of the mobile device is important, with the goal to prolong the usage time of the device. Thus, a small increase in response time is worth a larger decrease in power consumption.

Lastly, the adaptive fidelity technique is capable of effectively managing the resources of a user's mobile device. In a service-oriented pervasive environment, it is foreseeable that many applications will compete for the limited resources on the mobile devices, resulting in high utilization which will degrade the performance of the individual applications. The presented technique for adaptive fidelity monitors the resource usage, and adjusts different fidelity factors of each application according to the current resource constraints and predefined user preferences. When a resource such as bandwidth becomes constrained, applications will lower their dependence on this resource by adjusting appropriate fidelity factors, such as the resolution of received images. By predefining preferences to each fidelity factor, the user will not be distracted with on-the-fly decisions to increase or decrease the load on resources in order to better the performance; these decisions will be made behind the scenes, transparent to the user. This solution is fitting for a pervasive environment, since it prevents the user from being distracted by an over-utilized system, and also shifts the decision-making responsibility of adjusting fidelity factors away from the user and onto the system.

## 3 Security

The concept of an environment saturated with computing agents that seamlessly interact with one another as well as with human users lends itself to many potential exploitation hazards. Security measures will be required in order to guarantee the integrity of operations, however, the authentication overhead must be limited to ensure seamless interaction. Two main identification issues have been recognized and addressed in the literature [11]:

1. **User identification**: the computing environment needs to be aware of the identity of each user in order to properly handle access to data and services. The smoothness of the identification protocols is particularly critical when synchronizing and transferring data across multiple platforms.
2. **Service identification**: users need a method to check that a particular service is trustworthy before engaging in its use. This prevents malicious services, which may be disguised as trusted services, from infecting a mobile client with malware or viruses. It also gives the user a sense of comfort and confidence when calling a service, such that the intended service is the one that is actually called.

Albeit they are both related to the common area of minimal-overhead security maintenance, these two tasks are very different in nature and need separate treatment. We will address strategies for user and service identification in the next two sections.

## 3.1 User identification

The field of user identification is one of the longest standing issues in the context of computing services, and many secure solutions have been elaborated. First, it is appropriate to distinguish between user *authentication*, where the user claims an identity and the system verifies the validity of that claim, and user *recognition*, where the system proactively identifies a user (usually based on physical features) without any action on the user's part. User authentication can be further divided into two categories: *knowledge-based* (i.e., the identity claim is validated on the basis of some information provided by the user, usually a password or a PIN) and *token-based* (i.e., the identity is confirmed through the possession of a token, which can be anything from digital keys to smart cards).

Quite obviously, user recognition is conceptually more appropriate to the pervasive idea of a smart environment than user authentication, as the user will not need to undertake any distracting action to gain access to desired services. However, recognition is computationally more challenging than authentication, in that the system will have to compare the user's features against a potentially large database of identities in order to determine if the presented features match any stored entry. Authentication, on the other hand, only requires the comparison of presented information (i.e., username and password) against a single database entry in order to determine if the presented password matches that of the stored password for the presented username. This distinction implies that the relative efficiency of authentication with respect to recognition can only be determined on a case-by-case basis, establishing which of the two (the user-initiated process of authentication, or the waiting time to search a database in recognition) is more bothersome for the user. In the next two sections, we will examine two user identification methods: token-based authentication, and biometric recognition. We will compare and contrast the two approaches and describe the respective benefits and limitations.

### 3.1.1 Token-based user authentication: Kerberos protocol with smartcards

In the effort to strike an optimal balance between low computational complexity (and therefore transparency) and security of operations, secure hardware has promising features. It provides secure storage for highly sensitive information (such as the user's cryptographic keys), and also offers the ability to perform cryptographic operations in hardware, thereby allowing for efficient and transparent authentication strategies. The user is not responsible for the initiation and execution of authentication protocols, since the secure hardware contains all the necessary information to complete the authentication in an independent manner.

One possible strategy that has been proposed involves the integration of Kerberos V5 with a smartcard [10]. In order to understand the role played by the secure hardware, let us first examine an ordinary Kerberos authentication:

1. As a preliminary step, all users and services need to have a long-term cryptographic key registered with an Authentication Server.
2. When a registered user wants to login to a particular service, it will send the Authentication Server a request for an initial ticket to connect to the Ticket Granting Service. The Authentication Server will then provide a session key encrypted with user's long-term key.
3. The user will receive the session key, decrypt it and use it to request a ticket from the Ticket Granting Service (TGS) in order to connect to the desired service. The Ticket Granting Service will now return a session key encrypted with the initial Ticket Granting Service key, rather than the user's long-term key.

Kerberos thus provides two layers of indirection in order to limit the use of the user's long-term key, therefore increasing the security of this protocol. Short-term session keys are used to encrypt the communications between the user and service provider. Due to the volume of transactions encrypted with a short-term, these keys are susceptible to be being discovered by an eavesdropper. Yet, a discovered key only allows the eavesdropper to listen to a single session, as these keys change session-by-session. The long-term key is used in very few transactions, preserving its security. However, two issues remain unaddressed:

1. The long-term key will have to be used occasionally, since the session keys from the TGS are only temporary. Furthermore, the encryption/decryption operations will be performed on the user's workstation, which is not necessarily a secure place.
2. In order to guarantee a sufficient level of security, the long-term key will have to be protected by a safe password, which the user will have to enter every time the key is required. In a pervasive scenario, this might result in an excessive burden on the user.

In [10], an authentication method is used where the user's long-term key is stored in a smartcard. All the encryptions and decryptions are performed by the card, which guarantees the security of operations (addressing the first problem) and eliminates the need for a user password (addressing the second problem), in that all the information necessary to login is securely stored in the card.

Notice that this process will of course still suffer from the drawbacks of token-based authentication methods; above all is the risk connected to the token's misplacement. Whoever posses the smartcard can authenticate as the smartcard’s owner. However, simply introducing PIN-activated smartcards, at the expense of a little user interaction, might ease the problem.

### 3.1.2 User recognition: biometric applications

The basic working principle of biometric recognition can be summarized as follows:

1. A photometric sensor detects the user and converts the relative information into a digital form.
2. The digital information is then processed by a feature extractor, which isolates the traits which are relevant to the recognition process.
3. The user's features are then compared against a template database. The result of the comparison will return the user's identity or a no-match.

The first step of biometric recognition begs the question: which human characteristics should be used in the detection process? Ideally, these characteristics will have to satisfy two conditions: (i) the trait will have to be unique and non-reproducible, in order to unambiguously identify a user and avoid system circumvention and (ii) the information should be easy to acquire and acceptable on the user's end, since a technology that is invasive (like a retinal pattern read) or has criminal reminiscences (such as fingerprints) may be unwelcome to users and may disrupt an otherwise seamless environment.

There are currently several biometric sensors on the market and under development. Biometric sensors can be used to recognize facial features (using optical or thermal techniques), as well as a human iris, written signatures, fingerprints, and hand geometry. Facial thermography proves to be perfect solution, for it satisfies the two aforementioned requirements, as it is forgery-resistant and least invasive. A thermographic sensor will detect and map the temperature distribution on the user's face. Since the blood flow pattern seems to be unique to each individual (and also hardly modifiable through surgery), the identification is highly secure. The features are also relatively easy to acquire.

The second and third steps of biometric recognition bring about a different aspect of identification, namely the computational cost and performance. As mentioned earlier, one of the fundamental requirements of a pervasive environment is the transparency of operations; not only should tasks be performed with minimal (if any) user intervention, but their execution should also be as invisible as possible. The process of matching biometric features against a template should therefore carry a computational complexity that does not exceed a small fraction of the system's resources.

As a last remark, notice that security and invisibility are two conflicting objectives, in that, increasing one will necessarily degrade the other. This is well expressed in Figure 10, which shows the False Match Rate (FMR) - False Nonmatch Rate (FNR) plane. FMR represents the probability of erroneously detecting a match (i.e., when the matching procedure is not restrictive enough, due to computational limitations such as the pervasive requirement for invisibility), while FNR represents the probability of failing to detect a match (i.e., when the matching procedure is too conservative, as in situations where security is of utmost importance).

Different applications will result in different tradeoffs between FMR (computational simplicity) and FNR (security). A high security access application will have very restrictive matching rules, whereas applications where missing a match is undesirable (like criminal identification and forensics) will have a high FMR at the expense of FNR. Finding a pervasive environment's collocation on the graph is an open problem that requires careful examination of the computing capabilities of the system, the sensitivity of data and applications in the system's domain, and the details of the identification process (complexity of feature extractor, size of user template, and so forth).

When compared to the solution provided by secure hardware, biometric recognition offers a solution that is more transparent and resistant to forgery. At least for the type of identification described in this section, the recognition process is completely invisible and hassle-free. However, the present level of the biometric technology is far from the ideal, exploitation-proof condition that is required for a secure application. Feature extractors will also play a fundamental role in determining how secure and computationally expensive the process will be.

On the other hand, smartcards do not suffer from any of the previous issues related to pattern recognition. The identification is performed in the traditional, well-established cryptographic key methods, which do not suffer from FMR and FNR issues. However, the use of smartcards is less invisible than biometric recognition and will always be subject to the risk of misplaced or stolen cards, even in the case of password protected cards. While the risk of false identification with biometric systems is likely to be reduced by the constant improving technology, the risk of false impersonation due to fraudulent card possession will be very hard to control.

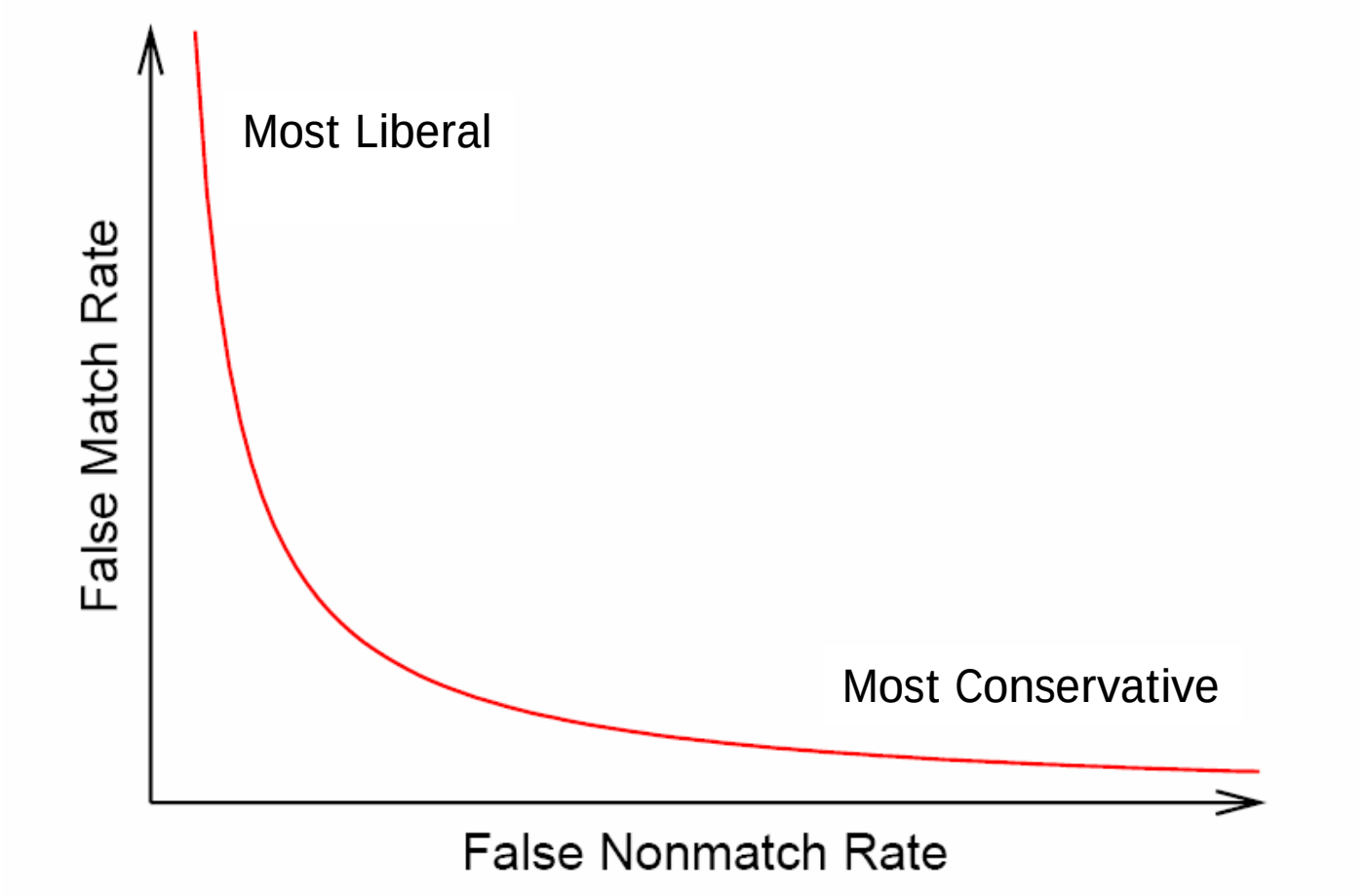

Figure 10 – Optimal set in the False Match Rate - False Nonmatch Rate plane. These two objectives, which represent respectively the computational simplicity and the security associated with a recognition method, are conflicting, and cannot be maximized simultaneously. An appropriate tradeoff between FMR and FNR depend on the goals and requirements of each specific application (image adapted from [11]).

## 3.2 Service Identification

A similar, yet distinct, security issue that is inherent to pervasive systems is the integrity of the services available to the user. In the pervasive scenario, the user does not initiate jobs; instead the jobs are proactively launched based on user preferences, history, and context. Therefore, it makes sense to provide service identification in addition to user identification. In other words, the system will have to be endowed with an authenticating structure that filters out possible fraudulent services and restricts job initiation capabilities and data access only to those services that have been authorized.

Similar issues are already present in the area of network systems, where users are constantly faced with the task of locating services (such as printing or storage services), verifying the trustworthiness of these services, and submitting a query to those trusted services. Many different methods for service lookup and discovery have been proposed in the past [13] [15], with philosophies and designs that are fitting to the pervasive idea. Such architectures usually involve the existence of discovery and authentication servers, which mediate the interaction between users and services by providing information about service capability, service authentication, and user authentication. In most cases, they will invoke the participation of other actors, such as additional certificate databases that physically hold the required information. The structure helps users search for particular services and verify their identity. At the same time, services will broadcast their presence and capabilities within the network, with the possibility of filtering (or customizing) the set of amenities offered to each user.

Similar types of network architecture, presently used to manage and secure the use of distributed computational resources in the standard sense (users initiating processes), can finally be adapted to pervasive system by letting the services, rather than the users, control the execution of tasks. The missing link necessary for the last step is represented by the integration of intelligent devices into the above architectures, i.e. the capability for storing the user's profile and performing pattern recognition on the present context. In the following section, we will discuss the secure Ninja Service Discovery Service (SDS) architecture [13].

### 3.2.1 Secure SDS with Ninja

The Ninja architecture is based on five classes of agents:

1. The users (clients)
2. The services
3. The SDS servers
4. The Capability Managers
5. The Certificate Authorities

Each SDS server governs its own domain, which is comprised of a number of agents from the other four classes. Domains are organized in a hierarchical fashion, allowing for easy scalability in case of overload; whenever a SDS server is unable to handle all the services and broadcasts in its domain, it will start a new child SDS server that will take over a sub domain. The following shows the interaction between the main entities:

- Each SDS server sends authenticated messages over a global multicast channel, which includes a description of the domain itself, descriptions of services registered in the domain, the multicast group address for each service in the domain, the address of the Certificate Authority, and the address of Capability Manager.
- Each service is responsible for registering itself with a live SDS server (who multicasts their presence via periodic advertisements). This implies that when a SDS server crashes, a service under this server is responsible for registering itself with a new server; this ensures some degree of fault tolerance. A service periodically multicasts its service descriptions (using authenticated and encrypted messages) to its SDS server as well as clients in its multicast group.
- Each client submits its queries for services to the SDS server responsible for its domain. The server will then reply with a list of matches corresponding to the client’s query, the available resources, and the user's privileges. It is assumed that there is trust between the clients and the SDS server, such that, the client can trust that the service descriptions received by the SDS server are accurate. However, this does *not* imply anything about the functionality or correctness of the services; it simply verifies the descriptions of the registered services. The client can then choose a service from list, and join the multicast group on which this service is being transmitted.

The core of the security framework of SDS is contained in the encryption of all messages between the system's entities, especially between servers and services. The use of asymmetric encryption would be the best choice for all encryptions, but efficiency requirements (which play an important role in simple network systems, and even more so in pervasive environments) suggest that a hybrid symmetric/asymmetric method would be best. Therefore, the service multicasts follow the three-segment format illustrated in Figure 11. The first part of the message contains the sender ID. The second part, ciphered with the SDS server's public key, contains several pieces of information (again the sender ID, the destination, etc.) along with a symmetric key that can be used to decipher the third, and largest, portion of the message, which is the actual payload. Thus, computationally expensive public-key decryption is only necessary to obtain the symmetric key, while computationally cheaper symmetric-key decryption can be done on the larger payload. This reduces the decryption overhead while at the same time securing the messages against eavesdropping.

| | | |
|---|---|---|
| Sender ID | Ciphered Text (containing symmetric key) | Payload (encrypted with symmetric key) |

Figure 11 – The format of a service broadcast

In addition to encryption, the system implements a global authentication procedure to guarantee the integrity of the associations between the system components and their public keys. In other words, security against fraudulent identities must be guaranteed not only by encrypted communications, but also through authentication of the endpoints. This is the role of the Certificate Authority and is accomplished in two steps:

1. The Certificate Authority collects certificates from the various system components.
2. Clients can query the Certificate Authority for a certificate to assess the validity of a public key associated with a service.

Since the keys and the certificates are public, the service of a Certificate Authority would not require computationally expensive encryptions when in operation (as the Certificate Authority only performs the signing of certificates offline), and therefore would blend well with a pervasive system where thin agents with low latency responses are a priority. The only requirement for implementing a Certificate Authority is that it must reside on a secure server.

The last component in this architecture is the Capability Manager, which stores the lists of clients' privileges in order to determine which user has access to which services. This greatly simplifies the amount of user interaction needed for each single query, since the SDS servers will prompt the Capability Manager for possible access restrictions, instead of asking the user to authenticate. Additionally, the user is only returned a list of matches which he or she is authorized to use; all other services are effectively invisible to the user.

Let us consider the example structure in Figure 12 depicting an SDS that manages resources in a computer science building. The hierarchy comprises four SDS servers, which can either be responsible for the administration of a specific physical location (such as the "CS Hall'", "4th Floor" and "Room 443" servers) or for the control of certain services (like the "Systems" server). In this figure, solid lines represent one-time communications, while dashed lines represent the system's various periodic broadcasts. One-time communications include the Remote Method Invocations by which each server can generate additional servers, as well as the clients' queries for services. The periodic broadcasts include the periodic server broadcast that disseminate the information about the server domain and the service broadcasts that publicize the available facilities.

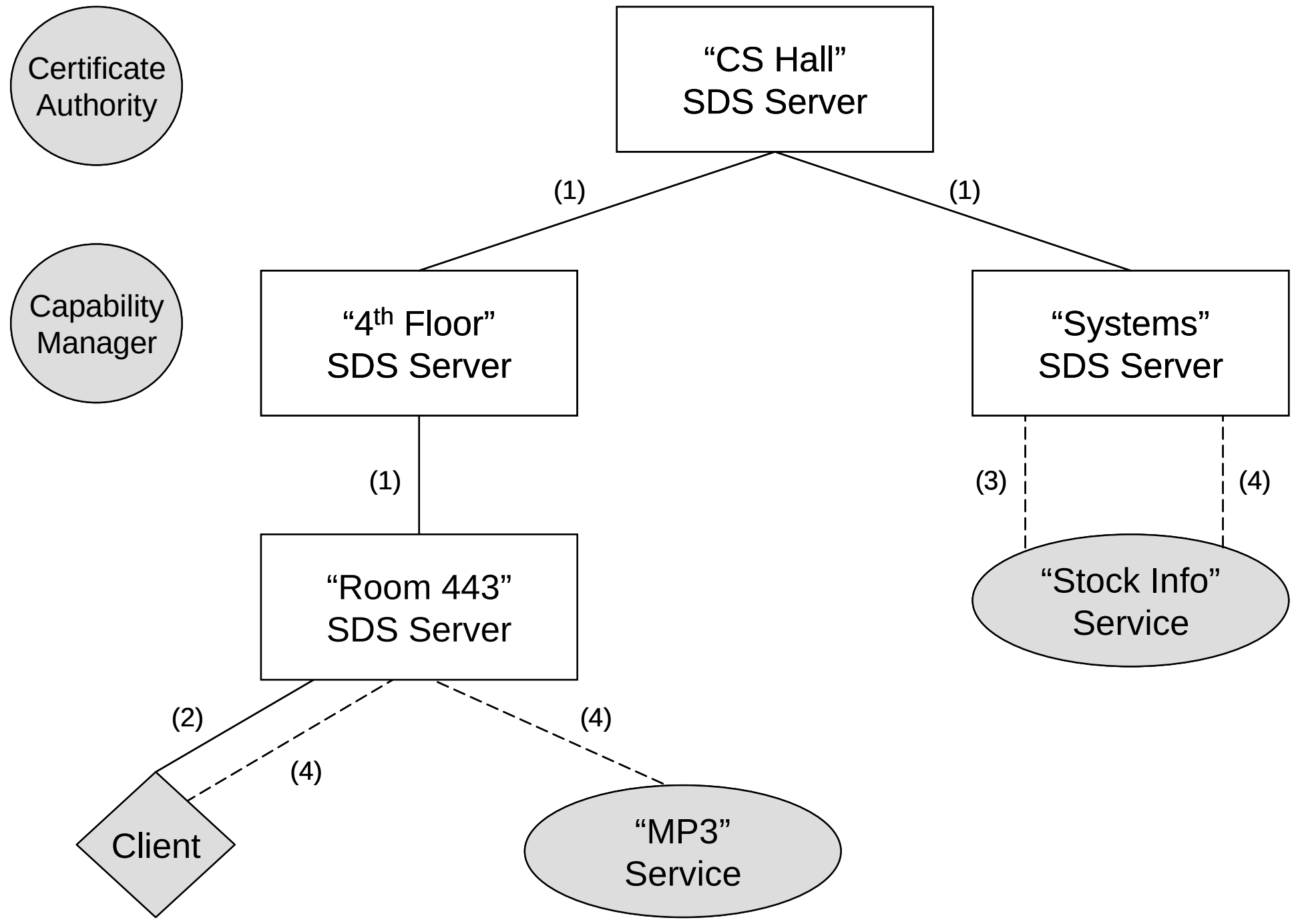

Figure 12 – The structure of a SDS architecture: dashed line indicate periodic broadcasts, while solid ones represent one-time communications. Lines marked with (1) are the authenticated server connections between a server and its offspring; lines marked with (2) represent the authenticated client connection; lines marked with (3) are the service broadcasts and lines marked with (4) are the server broadcasts (image adapted from [13]).

From a pervasive perspective, this architecture provides an infrastructure that satisfies the demand for a transparent, yet trustable, provisioning of services. The identification of services is assigned to the SDS servers, which keep track of their identity (which has been verified by a Certificate Authority), and only share with the clients the existence of those services with trusted identities. All a client needs to do is submit a service query to the server, and the resulting matches will be, by construction, already authenticated. As an aside, the structure provides the additional benefit of a straightforward organization of the user's privileges through the Capability Manager. Lastly, it should be strongly noted that the level of trust guaranteed by this architecture is only as strong as the trust between a client and the Certificate Authority, and the trust between a client and the SDS server.

In order to incorporate this architecture into a pervasive system, an additional step is required to eliminate the need for explicit queries for services made by the user, transferring this task to the control of the mobile device itself.

## 3.3 Conclusions

This section has presented a set of available tools and platforms to address security issues in the specific context of a pervasive environment, i.e. in a system with strong constraints on transparency and minimal user distraction. The issues of identification have been separated into two categories: user identification and service identification. Furthermore, user identification has been partitioned into authentication and recognition, and each has been reviewed in order to offer a spectrum of different (and possibly, complementary) identification mechanisms which realize different degrees of security and transparency.

Finally, we have described a model service identification architecture equipped with a selective discovery protocol that not only prevents unauthorized users from using the available resources, but also prevents them from detecting those services to which they have not been granted access, increasing both the efficiency and the security of the system in a context-adaptive fashion.

# 4 Current Projects

As a realistic illustration of the challenges posed by resource management and security requirements, as well as the practical solutions that have been devised over the years, we present here a few projects that are currently under way.

The applications of pervasive computing span the entire spectrum from generic, multipurpose distributed computing to the implementation of specific services like home automation or pervasive healthcare. An extensive list of ongoing projects is shown in Table 2. This section will focus on a select group of these projects, including:

- Multipurpose systems providing proactive services to the users: MIT's Oxygen, Carnegie Mellon's Project Aura, and Berkeley University's Smart Dust all belong to this category.
- The Pervasive Continuous Curriculum (PCC) project from Pennsylvania State University [14] and AULA from University of Castilla investigate the application of pervasive systems to academic issues, such as the administration of classes and the design of individual curricula.
- Medical services: the projects MyMD (from MIT) and TMBP (Centre for Pervasive Healthcare, Denmark) provide a framework for healthcare monitoring.
- Smart environments naturally lend themselves to home, office, and urban automation applications. Existing projects propose the implementation of pervasive devices at several different scales, spanning from the automation of daily home tasks (like powering on/off the lights or operating appliances) to the concept of connectivity and service provision as integral components of urban design and architecture.

Table 2 – A sample of current pervasive computing projects, by category

| MULTIPURPOSE SYSTEMS |
|---|
| Oxygen (MIT, [16])<br>Aura (CMU, [17])<br>(Berkeley, [18])<br>Spectacles (Johannes Kepler Universität Linz, [19])<br>PerComp (Federal University of Campina Grande, [20])<br>Application SuperNetworking - All-IP (University of Oulu, [21])<br>Particles (University of Munich, [22])<br>Mundo (Technische Universität Darmstadt, [23])<br>MOBIUS (European Mobius consortium, [24])<br>OTOGI (Waseda University, [25])<br>ACAMUS (Kyung Hee University, [26])<br>LOCAL (University of Minho, [27])<br>Disappearing Computer Initiative ([28]) |
| **EDUCATION** |
| PCC (PSU, [14])<br>AULA-IE (University of Castilla - la Mancha, [29]) |
| **HEALTHCARE** |
| MyMD (MIT, [30])<br>Context Aware Health Monitoring (University of Technology, [31])<br>Abaris (Georgia Institute of Technology, [32])<br>TMBP (Denmark Centre for Pervasive Healthcare, [33])<br>Centre for Pervasive Healthcare, [33])<br>UniCare (Imperial College, [34]) |
| **HOME, OFFICE AND URBAN AUTOMATION** |
| LiveSpaces (University of South Australia, [37])<br>FlexHaus (Fraunhofer Institut SIT, [38])<br>SSLab (Keio University, [35])<br>SmartLab (University of Deusto, [36])<br>Interactive Workspaces (Stanford University, [39])<br>Cityware (Imperial College et al., [40])<br>Shared Worlds (University of Limerick, [41]) |

## 4.1 Multipurpose Systems

One of the founding ideas of pervasive systems is the concept of a single user served by a multitude of computing agents, which saturate the environment in order to monitor the context, predict the user's intent and offer a wide range of services (ideally, all services that are necessary to the user and do not require his or her interactive participation) in a proactive fashion. A vast number of current pervasive projects deliver a generic framework for the provision of services, from a simple weather forecast update to more complex business transactions. In this section, we will describe two such frameworks: MIT's Oxygen and Project Aura from Carnegie Mellon University.

### 4.1.1 Transparent security with Oxygen

Traditionally, the interaction between humans and computers has required humans to learn and adapt to the logic and working principles of the specific machinery at hand. Conversely, the Oxygen project orbits around the idea of bridging this interaction by teaching computers to communicate in a human-friendly manner, supplying the user with service-providing agents that completely mask the underlying technology. This framework involves three principal entities: Users, Devices, and Networks:

- **Users:** the human clients, and main focus, of the system. In other words, the computational platform provides a set of technologies that enable the user to automate tasks, network with other users, and communicate information in a completely natural and transparent fashion:
    - *Automation:* low-level actions are represented by basic automation objects, and user technologies include scripting tools capable of manipulating these objects and constructing arbitrarily sophisticated actions from the basic building blocks. The objects can be *physical* (which can include perceptual devices, temperature or light sensors, and power switches among others) or *virtual* (which comprise software agents and daemons capable of processing information and making decisions). As an example, a set of physical objects could be combined in a script specifying user preferences such as indoor temperature, light and sound levels, computer screen resolution, preferred font size and so forth; this script could then be run whenever the user enters a building or a computer lab, allowing him or her to concentrate on high level tasks rather than on adjusting the environment settings.
    - *Collaboration:* the system also provides a platform that keeps track of the interactions between users, using context-aware agents to classify the content, the properties, and the parties involved in each specific collaboration instance. This information is then passed to the global system to provide collaboration-related services, such as teleconference infrastructure, scheduling of meetings, and collaboration in database management.
    - *Knowledge access:* data can be produced, searched and shared among users with the aid of the knowledge access subsystem, equipped with semantic search capabilities and tools for extensible data representation and acquisition. The Haystack platform

[42], the Semantic Web [43] and the START language [44] are all components of this subsystem.

- **Devices:** the entities responsible for detecting the user's intent and providing the appropriate services. They can be portable or embedded in the environment.
  - Users are provided with Handy21s (H21s), i.e. handheld devices that are associated with a specific user, rather than with an environment. The handhelds provide continuous connectivity for the users wherever they are, at any time.
  - Vehicles, buildings, and public spaces are assigned one or more computing agents called Enviro21s (E21s). These are responsible for the provision of services that pertain to the given environment, such as receiving and sorting telephone calls within a certain building or delivering travel information and driving directions in a user's car.

  A visual representation of these two categories is illustrated in Figure 13, showing the implementation of devices in a given environment: the space is pervaded by several Enviro21s (for instance, one in each room of a building); additionally, the mobile devices Handy21s move across the domain communicating with the relevant Enviro21s and between each other, as appropriate.
- **Networks:** the infrastructure that establishes connectivity between users and devices. The networks (N21s) can dynamically reconfigure themselves to provide the interconnection between subsets of computers, also referred to as *collaborative regions*. Networks also play a fundamental role in the discovery of services by mobile agents. When a user enters a new smart space, the user's handheld device will automatically explore the surrounding environment, through the provided network, and record the available services for future use.

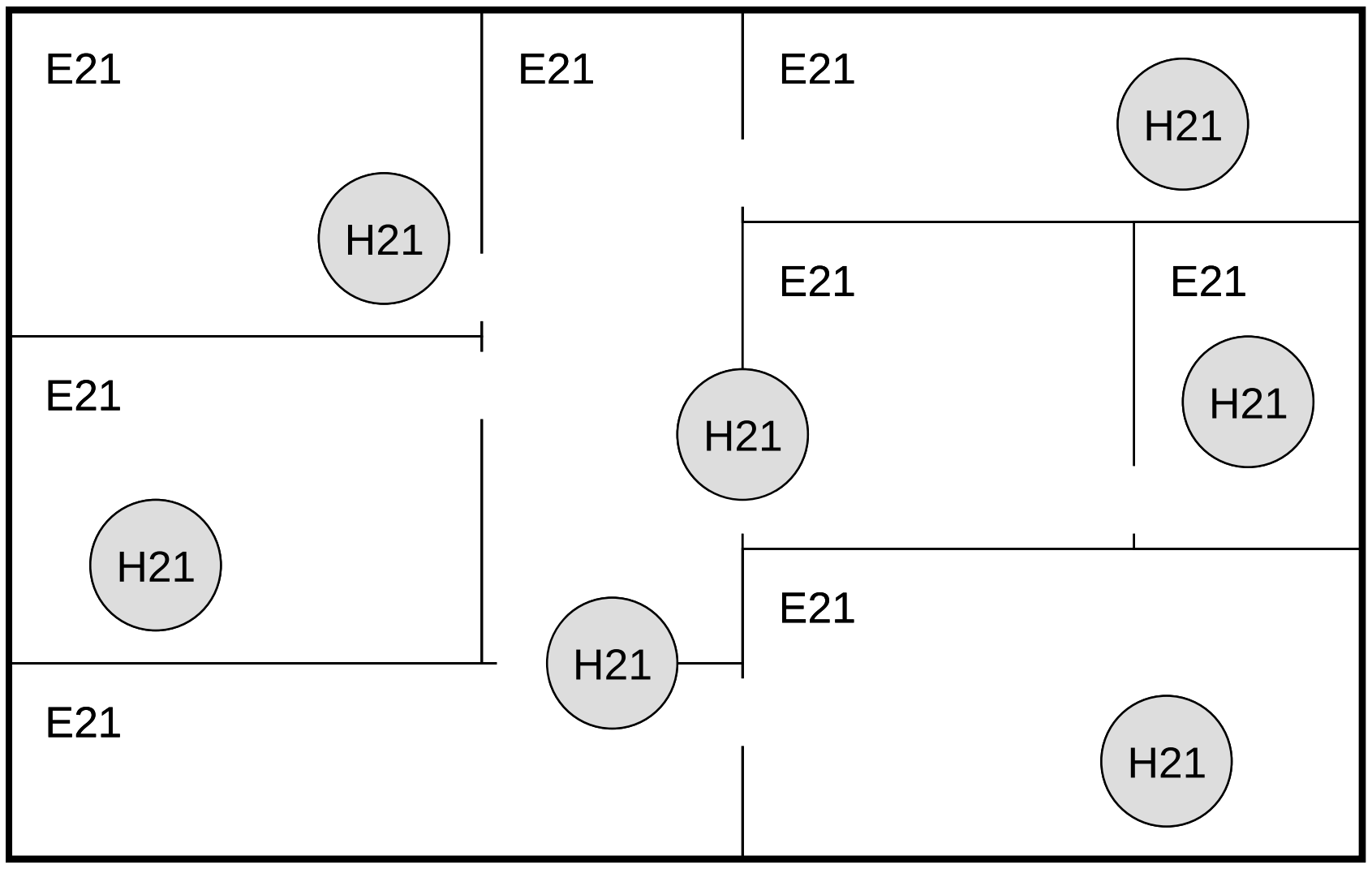

Figure 13 – A typical Oxygen environment, with fixed devices embedded in each section (for instance, each room of a building) for the provision of room-related services such as controlling the appliances, guaranteeing the access to a certain set of users only, etc. Handheld devices move through this structure communicating with the ambient devices and between each other, and negotiating all the low level procedures that are required by each action, thereby absorbing most sources of user distraction.

Oxygen thus provides a realization of several of the security techniques described in Section 3. For ordinary operations, the system resorts to a token-based method of authentication. The tokens are, in this case, the handheld devices H21s, which are furnished with the capabilities to authenticate the users with the surrounding services, eliminating the need for any other type of access control (the system is also endowed with a Discovery Service architecture responsible for resolving the service requests). Furthermore, the token-based authentication can be combined with more secure identification methods (such as fingerprint identification implemented on the H21s) for those specific applications that require an increased level of protection, such as bank transactions or access to sensitive information. Additionally, the H21s can be instructed to identify each other, in order to form a collaborative region, i.e. a self-organized set of mutually authenticated users who may share data or have access to specific services. The routing infrastructure is provided by Chord [45], a scalable framework for peer-to-peer overlay networks.

#### 4.1.2 Project Aura: resource management by a user proxy

Project Aura from Carnegie Mellon University is a pervasive platform based on the concept of a personal *aura*, i.e. an abstract representation of the user's intent and preferences, which facilitates the user's mobility by taking control of all the migration-related duties [17].

In particular, the project addresses four types of causes of user distraction due to heterogeneous computing infrastructure: (i) a migration to a different environment, as the user moves between different locations and chooses to transfer the pending tasks across devices; (ii) a change in available resources, as network connectivity and/or computing power fluctuate due to time-varying system load; (iii) a change in the executing task, due to a change in the user's task priority; and (iv) a change in the context, which requires the suspension of prior tasks and the inception of new ones.

Ideally, the user proxy plays the role of a coordinating entity that decides on the services to request, on the quality-of-service that can be considered acceptable, and on all the other issues related to the aforementioned distraction sources. The system architecture comprises four components:

- The **Task Manager**, or **Prism**, constitutes the user proxy. Prism resorts to a platform independent description of tasks, treated as high-level, conceptual objects rather than just as a collection of specific applications. This allows the system to be more aware of user intent (e.g., denoting one's action as “editing a text document" as opposed to “using Microsoft Word") and to easily migrate the tasks between different platforms.
- The **Service Suppliers** provide the service wrappers for the specific environment they reside in. The Suppliers will respond to an abstract service request (such as “editing a text document") by invoking the specific application (e.g., Microsoft Word) present in the current environment. This is the system component that targets the heterogeneity of computing environments by encapsulating all applications with the required capabilities in one high level wrapper. The other system components do not need to be aware of these details.
- The **Context Observer** monitors the physical context, passing the relevant information to the other units in charge of the migration of tasks. The Context Observer is, for instance, responsible to detect when a user leaves a specific environment and joins a new one.

- The **Environment Manager** plays the role of a domain coordinator, monitoring the available resources and organizing the users' requests.

Every environment is equipped with an Environment Manager, a Context Observer, and an instance of Task Manager for each user in the environment. The environment also contains Suppliers for each available service, which is registered with the Environment Manager through a XML-based feature description system. A pictorial representation of the system interactions is provided in Figure 14.

Project Aura represents a solution to the resource management problem that relies on a complex unit (the Prism) making smart decisions based on context and user intent.

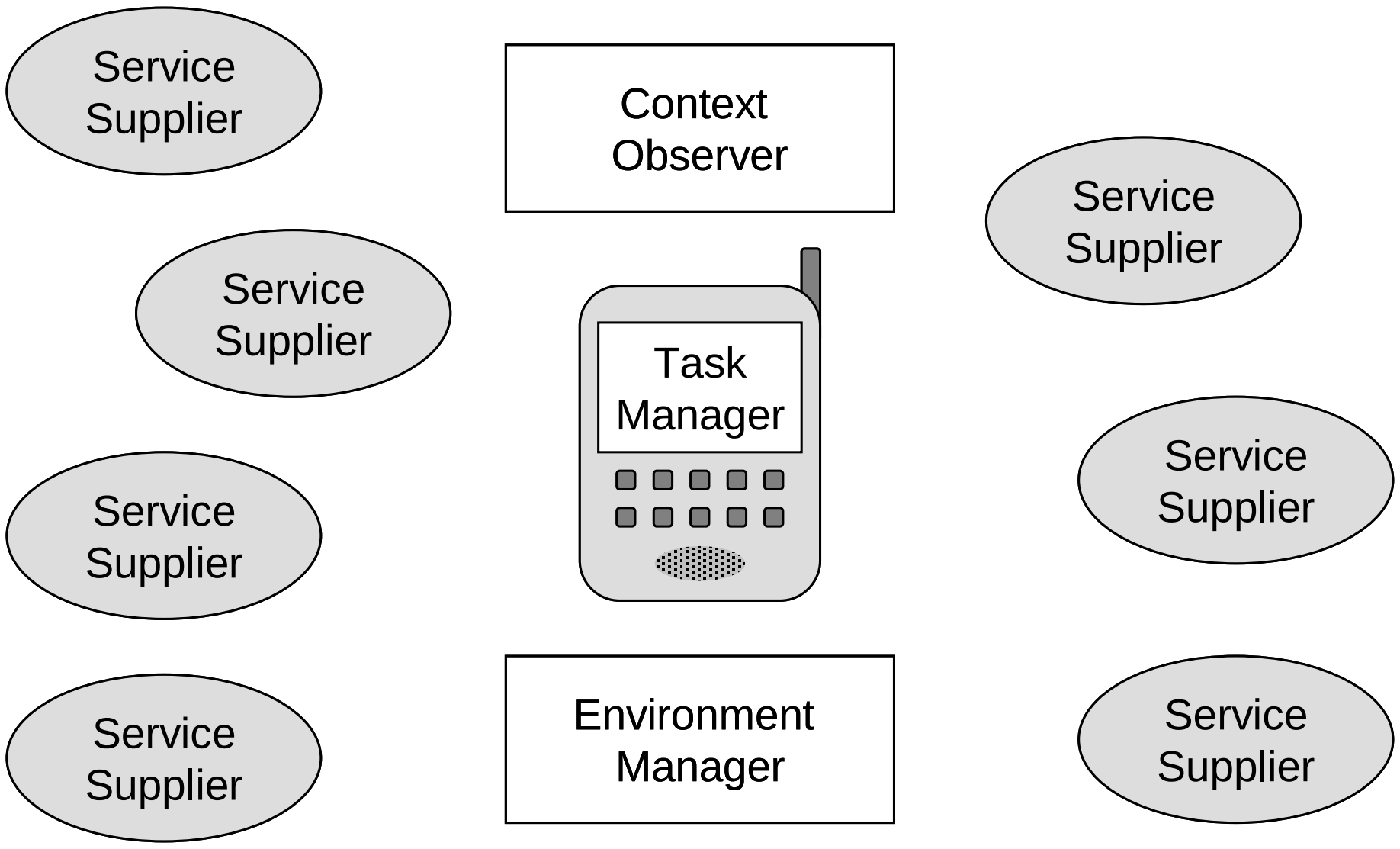

Figure 14 – The outline of Aura's architecture

## 4.2 Dedicated Systems

The concept of computing agents that permeate the environment has also stimulated work in the realm of specific applications, such as education, healthcare, unmanned vehicles, and home automation.

Quite understandably, predicting user intent and preferences in a dedicated environment is simpler than it would be in a generic environment, as the system architecture can be tailored to the specific properties and constraints of the application at hand. In other words, it is feasible to trade some of the flexibility of multipurpose systems for simple and accurate prediction of user intent. We will now describe two application-driven scenarios, illustrating how some of the issues of non-specialized frameworks find here a natural solution.

### 4.2.1 Pervasive education

The task of curriculum design in collegiate programs is an elaborate, often time-demanding procedure that involves evaluating the student interests, establishing the required *corpus* of knowledge or expertise needed for a specific degree and avoiding, to the largest possible extent, duplication of material across courses.

There is a conspicuous list of issues connected with such a task:

- **Curriculum development and access tools:**
  - To each degree, it is necessary to associate a set of modules that constitute the essential, indispensable core of knowledge for that specific curriculum. This set should also be as uniform as possible across different institutions, in order to guarantee that the same denomination is indeed affixed to the same educational path; this is particularly suited to computer science and engineering degrees, but can be extended to other areas.
  - On top of the standardized core, each student can build a collection of individual subfields of expertise. The selection of the appropriate units involves the interaction between the student, his or her academic advisor, and the faculty members offering course modules in related areas. This process presupposes the existence and availability of relevant, up-to-date information, and the ability to process this information efficiently.
- **Teaching practices:** The course offering itself should present a similar degree of flexibility and 1-1 interaction between parts. In the first place, class syllabi should be fine-grained to allow individual students to only select the topics which they have not yet learned. Ideally, the granularity should eventually be so fine that the degree consists in a continuous knowledge acquisition rather than a discrete course-based learning process. Secondly, interactive methods should both facilitate the course offering and increase classroom participation, making sure the student's progress is fed back directly into the system.

It is apparent that the advances obtained in the area of database management, learning structures, distributed computing, mobile agents and pervasive systems provide a feasible solution to many of these issues. As an example, the Pervasive Continuous Curriculum (PCC) project constitutes an effort to collect the relevant technologies emerging in these fields in order to construct a pervasive education framework. The platform includes three sets of components:

- The set of instructors, $I$
- The set of students, $S$
- The set of courses, $C$.

The members of each set and their interactions are realized on the backbone provided by the Pervasive Information Community Organization (PICO) framework [14], which consists of software agents (called Intelligent Delegates or *delegents*) that can self-organize into dynamic communities with the purpose of sharing data between one another, processing different sources of information, and making context-aware decisions. For example, as a course is scheduled within an academic program, a course delegent $D_c$ is created, which holds information about the course syllabus and records the student delegents $D_s$ that are created for each student that registers for the course. The $D_s$ will perform extensive checks on each student's background to determine whether he or she meets the prerequisites to attend the class and whether the class contents match the student interests and/or satisfy the chosen degree requirements. The $D_s$ will also interact with the instructor's delegent $D_i$ to create an effective 1-1 learning scenario, where individual questions and difficulties are addressed on an adaptive, personalized basis. An illustration of the system's components and interaction is shown in Figure 15.

### 4.2.2 Pervasive healthcare

In this section, we will describe how a pervasive system can offer continuous healthcare monitoring for patients with critical medical conditions. Quite understandably, the delicate issues involved in patient treatment demand for a strict set of constraints regarding the system's quality-of-service and, to some extent, privacy and security.
One such system is MyMD, a project developed at MIT [30]. The platform is composed of five entities:

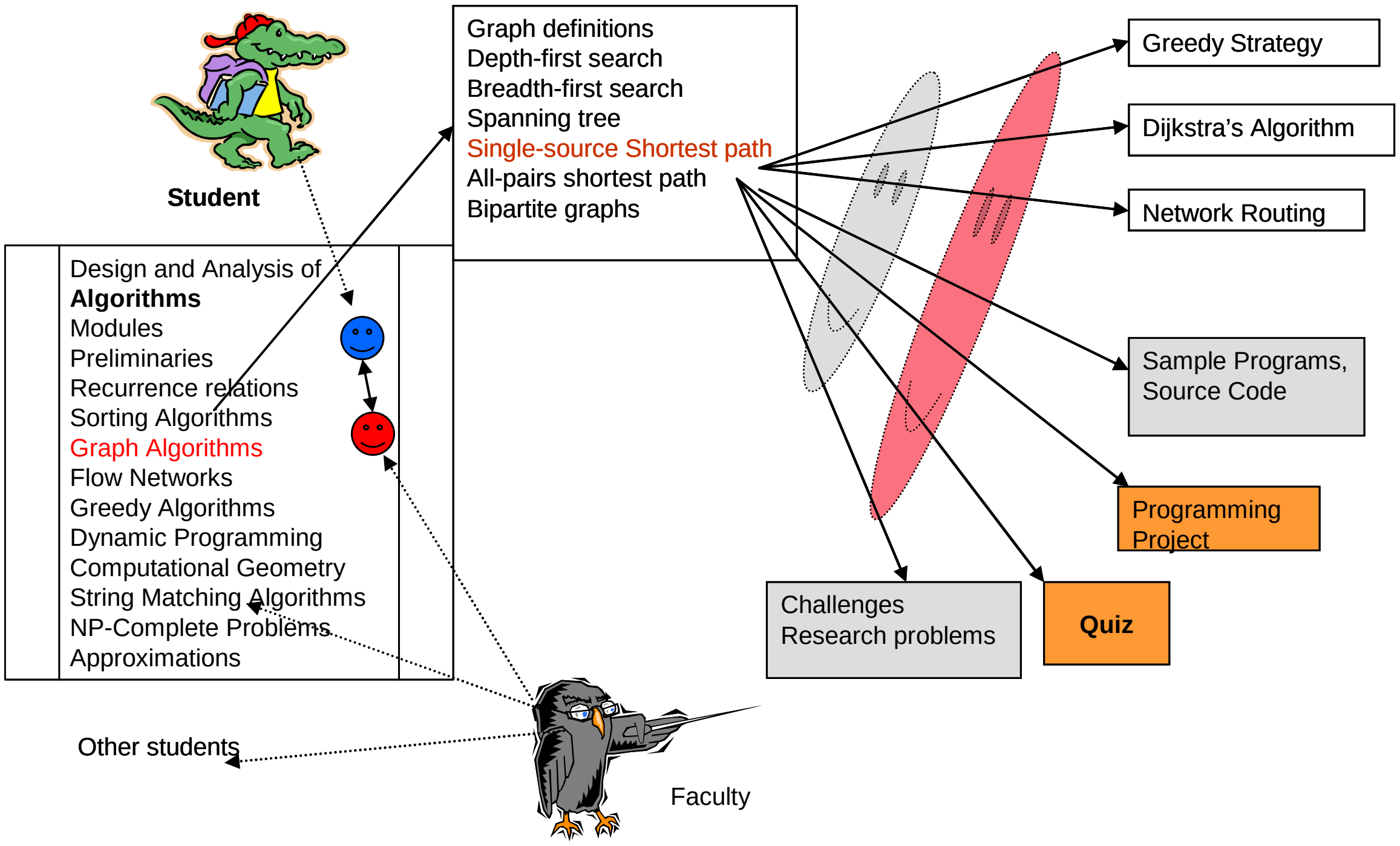

Figure 15 – A model of the interactions involved in course administration using a computer science course in algorithms as an example

- Sensors, which monitor the patient's vital signs
- Sensor proxies, in charge of combining and coordinating the readings of all the different sensors
- A real-time streaming database, for fast access to the patient's history (both for reading and updating purposes)
- Storage capabilities, to provide database support
- Communication facilities to promptly issue alerts and provide emergency directions

The realization of this architecture poses several challenges within the domain of resource management: the medical devices will need to be portable yet autonomous entities, capable of performing their task with minimum energy consumption. The system will also have to constantly, proactively monitor the network conditions in order to predict possible failures and promptly effectuate the transition to alternative procedures in case communication is lost.

## 4.3 Conclusions

In order to provide a realistic sample of the features and requirements for the implementation of a pervasive system, we have reviewed a sample of ongoing pervasive computing projects, along with their strategies to tackle the challenges described in Sections 2 and 3.

A crucial element in this scenario is represented by the following inherent challenge to pervasive systems: while dedicated applications come with intrinsic and well-defined notions of the minimum quality-of-service and security levels necessary for safe operation, generic platforms required by pervasive environments must be more flexible to satisfy the heterogeneous requirements of the clients, service providers, and intermediate networks.

# 5 Final Remarks

Pervasive computing is the third generation of computing in which many computing devices, of different shapes and forms, concurrently serve the individual user. Much of the groundwork for pervasive computing, in respect to hardware, is already present, such as wireless networks, powerful mobile devices and an abundance of workstations and servers. Furthermore, many traditional theories and algorithms are applicable to pervasive computing. Yet, there is an important property of pervasive computing which requires these traditional hardware and software elements to be modified; it is the requirement for a transparent integration between the user and the computing resources. Services must be performed for the user in a least-invasive, distraction-free manner. A possible resource bottleneck at the surrogate servers, who act as service providers to multitudes of mobile devices, may disturb the users' interaction in the pervasive environment. Additionally, the constant need to authenticate oneself with different services in the environment can obstruct the user's interaction. This chapter presented ways to smooth this interaction, and maintain a distraction-free environment.

Regarding resource management, the techniques of distributed caching, broadcasting, and adaptive fidelity were offered. Distributed caching and broadcasting alleviate the bottleneck at the service providers. Distributed caching spreads the workload across the mobile devices, while broadcasting eliminates the need to process redundant requests and transmit redundant responses. Both techniques scale well, and are therefore suitable for a large-scale pervasive environment. Since mobile clients become mobile servers with distributed caching, as more clients enter an environment, the increased client demand will be met by an increase in server availability. In addition, the spread of stale date in a caching environment can be thwarted by imposing some quality-of-service guarantee on the data received. In regards to broadcasting, the resources consumed in a broadcast are independent of the number of clients, thus implying the scalability of this technique. Moreover, the power of a mobile host can be conserved when an indexing scheme is integrated into a broadcast, and the response time for a request can be decreased as more parallel channels are introduced. Further, to efficiently utilize the limited resources of a mobile device, an adaptive fidelity technique can be implemented to dynamically adjust the fidelity of services in a manner which yields the greatest utility under the given set of resource constraints. The presented techniques for adaptive fidelity are suitable for pervasive computing as they offer portability and limited user distraction. Since resource consumption can be modeled empirically, regardless of the architecture and organization of a machine, this aspect of an adaptive fidelity algorithm can be ported to the diverse assortment of machines in a pervasive environment. By considering fidelity parameters independently from one another in respect to their offered utility, less setup is required from the user and less storage is required for these settings; thus, limiting the interaction from the user and limiting the storage requirement on the resource-constrained mobile device. With sufficient resources in the available service providers as well as in the mobile clients and communication network, these resource management techniques would not be necessary; yet for a realistic implementation of pervasive computing, resource management techniques should be employed to compensate for an under-provisioning of resources and to allow for a distraction-free environment.

We have also discussed how multiple security issues must be addressed before pervasive systems become a trustable environment to perform important tasks. In the first place, the identities of the system's clients need to be verified in a way that is both secure and transparent. While the integrity of the system operation is naturally a non-negligible concern, the pervasive philosophy brings about an additional, conflicting priority: to minimize distraction to the user, allowing his or her focus to remain on the high-level tasks. The design of valid identification strategies that integrate themselves into the seamless ensemble of pervasive components is an extremely complex task. Token-based authentication with smartcards and biometric recognition have been discussed and contrasted against each other in order to present a profile of the complications involved. In addition, the idea of a smart environment initiating processes on behalf of the user demands for an additional identity check: the services and the service providers themselves will have to be authenticated before they can be integrated into the system. We have discussed an architecture that provides service identification without placing any additional burden on the user's workload. Before a similar type of organization becomes effective in the pervasive context, the concept of users querying for services will have to be replaced with the concept of processes initiating services based on the situation and on the user's preferences and history. Drawing upon the field of artificial intelligence to supply the groundwork for context-aware devices will provide this last step towards the third generation of computing.

As a practical counterpart of our analysis, we have reviewed a few current attempts to achieve a safe yet transparent pervasive operation, commenting on the many different strategies and designs.

## 6 Acknowledgments

We wish to thank Robert Collins for providing helpful comments on this work, and Kevin Grady for carefully reviewing and revising this manuscript. This work in part has been supported by the National Science Foundation under the contracts IIS-0324835.

# 7 References

[1] M. Weiser, "The Computer for the 21st Century", *Scientific American*, 256(3), September 1991, pp. 94-104.

[2] F. Perich, A. Joshi, T. Finin, Y. Yesha, "On Data Management in Pervasive Computing Environments", *in IEEE Transactions on Knowledge and Data Engineering*, Vol. 16, No. 5, May 2004, pp. 621-634.

[3] T. Imielinski, S. Viswanathan, and B. R. Badrinath, "Data on Air: Organization and Access", *in IEEE Transactions on Knowledge and Data Engineering*, Vol. 9, No. 3, May-June 1997, pp. 353-372.

[4] T. Imielinski, S. Viswanathan, "Adaptive Wireless Information Systems", *in Proceedings of the Special Interest Group on DataBase Systems*, Japan, October 1994, pp. 19-41.

[5] D. Narayanan, J. Flinn, M. Satyanarayanan, "Using History to Improve Mobile Application Adaptation", *in Proceedings of the Third Workshop on Mobile Computing Systems and Applications*, Monterey, CA, December 2000, pp. 31-40.

[6] A.R. Hurson, Y. Jiao, and B. Shirazi, "Broadcasting a Means to Disseminate Public Data in a Wireless Environment: Issues and Solutions", *Advances in Computers*, Vol. 67, 2006, pp. 1-85.

[7] J. Sustersic and A.R. Hurson, "Quality of Service (QoS) in Internet Cache Coherence", *Journal of High Performance Computing and Networking*, Vol. 3, No. 5/6, 2005, pp. 296-308.

[8] V. Poladian, J. Sousa, D Garlan, M Shaw, "Dynamic Configuration of Resource-Aware Services", *in Proceeding of the 26$^{th}$ International Conference on Software Engineering*, May 2004, pp. 604-613.

[9] J. Sousa, D. Garlan, "Aura: An Architectural Framework for User Mobility in Ubiquitous Computing Environments", *in Proceeding of the Third IEEE/IFIP Conference on Software Architecture*, Vol. 224, Montreal, 2002, pp. 29-43.

[10] N. Itoi and P. Honeyman 1999, "Practical Security Systems with Smartcards", *in Proceedings of the Seventh Workshop on Hot Topics in Operating Systems*, Arizona, 1999, pp. 185-190.

[11] A. Jam, L. Hong and S. Pankanti 2000, "Biometric Identification", *in Communications of the ACM*, Vol. 43, Issue 2, 2000, pp. 90-98.

[12] M. Satyanarayanan 2001, “Pervasive Computing: Vision and Challenges”, *in IEEE Personal Communications*, Vol. 8, Issue 4, 2001, pp. 10-17.

[13] S.E. Czerwinski, B.Y. Zhao, T.D. Hodes, A.D. Joseph and R.H. Katz 1999, “An Architecture for a Secure Service Discovery Service”, *in Proceedings of the Fifth Annual ACM/IEEE International Conference on Mobile Computing and Networking*, 1999, pp. 24-35.

[14] A.R. Hurson, E. Jean, M. Ongtang, X. Gao, Y. Jiao and T.E. Potok, 2007, “Recent Advances in Mobile Agent-Oriented Applications”, *in Mobile Intelligence: When Computational Intelligence meets Mobile Paradigm*, L.T. Yang and A.B. Waluyo (editors), John Wiley & Sons.

[15] J. Mauro and G. Minden, 2004, “Security Model in the Ambient Computational Environment”, M. Sc. Thesis, University of Kansas.

[16] Project Oxygen’s website: oxygen.csail.mit.edu

[17] Project Aura’s website: www.cs.cmu.edu/Thura; J. Sousa, D. Garlan, “Aura: An Architectural Framework for User Mobility in Ubiquitous Computing Environments”, *in Proceeding of the Third IEEE/IFIP Conference on Software Architecture*, Vol. 224, Montreal, 2002, pp. 29-43.

[18] SmartDust’s website: robotics.eecs.berkeley.edu/pister/SmartDust

[19] Spectacles’ website: www.pervasive.jku. at/Research/Projects/SPECTACLES

[20] PerComp’s wiki: wiki.percomp.org

[21] All-IP’s website: www.mediateam.oulu.fi/projects/allip

[22] Particles’ website: particles.teco.edu

[23] www.tk.informatik.tu-darmstadt.de/Forschung/Poster/Mundo

[24] Mobius’ website: mobius.inria.fr

[25] Interaction Group’s website: www.dcl.info.waseda.ac.jp/groups/intg.html

[26] Acamus’ website: uclab.khu.ac.kr/camus

[27] LOCAL’s website: get.dsi.uminho.pt/local

[28] DC’s website: www.disappearing-computer.net

[29] AULA’s website: chico.Inf-cr.uclm.es/AULA_IE

[30] MyMD's website: mymd.csail.mit.edu

[31] www-staff.it.uts.edu.aut~peterl/mobilelab/research/project1_health.html

[32] Abaris' website: home.cc.gatech.edu/julie/24

[33] TMBP's website: www.tmbp.dk

[34] UbiCare's website: www-dse.doc.ic.ac.uk/Projects/ubicare

[35] SSLab website: www.ht.sfc.keio.ac.jp/SSLab

[36] SmartLab's website: www.smartlab.deusto.es

[37] University of South Australia e-World Lab page: e-world.unisa.edu.au

[38] www.sit.fhg de/_SITProjekte/flexhaus

[39] Interactive WorkSpaces' website: iwork.stanford.edu

[40] Cityware's website: www.cityware .org.uk

[41] Shared Worlds' website: www.shared-worlds.org

[42] Haystack's website: haystack.csail.mit.edu

[43] Semantic Web's page: www.w3.org/2OO1/sw

[44] START's website: start.csail.mit.edu

[45] Chord's website: pdos.csail.mit.edu/chord